\documentclass[preprint,nofootinbib,superscriptaddress,longbibliography]{revtex4-1}
\usepackage{amsmath,amssymb,bm,graphicx,xcolor,multirow}

\usepackage{subfig, braket}

\usepackage[
colorlinks=true, 
pdfstartview=FitV, 
linkcolor=blue, 
citecolor=magenta, 
urlcolor=blue, 
bookmarks=true,
bookmarksnumbered=true,
pdftitle={},
pdfauthor={}
]{hyperref}

\usepackage[normalem]{ulem}  % \sout{old text} for strikeout

\allowdisplaybreaks[1]

\newcommand{\+}{\dagger}
\newcommand{\ra}{\rightarrow} 
             
\newcommand{\CP}{{\mathbb C \mathbb P}}
\newcommand{\RP}{{\mathbb R \mathbb P}}
\newcommand{\Z}{{\mathbb Z}}

\newcommand{\U}{\mathrm{U}}
\newcommand{\SU}{\mathrm{SU}}
\newcommand{\PSU}{\mathrm{PSU}}
\newcommand{\SO}{\mathrm{SO}}

\newcommand{\I}{{\mathbb I}}

\newcommand{\mC}{{\mathcal C}}
\newcommand{\mD}{{\mathcal D}}
\newcommand{\mL}{{\mathcal L}}
\newcommand{\mP}{{\mathcal P}}
\newcommand{\mT}{{\mathcal T}}

\newcommand{\mR}{{\mathcal R}}

\newcommand{\mM}{{\mathcal M}}
\newcommand{\mN}{{\mathcal N}}

\newcommand{\diff}{\mathrm{d}}
\newcommand{\rme}{\mathrm{e}}
\newcommand{\rmi}{\mathrm{i}}

\newcommand{\tr}{{\rm tr}}

\begin{document}

\title{
Global anomaly matching in higher-dimensional $\CP^{N-1}$ model
}

\author{Takuya Furusawa}
\email{furusawa@stat.phys.titech.ac.jp}
\affiliation{Department of Physics, Tokyo Institute of Technology, Ookayama, Meguro, Tokyo 152-8551, Japan}
\affiliation{Condensed Matter Theory Laboratory, RIKEN, Wako, Saitama 351-0198, Japan}

\author{Masaru Hongo}
\email{masaru.hongo@riken.jp}
\affiliation{Department of Physics, University of Illinois, Chicago, IL 60607, USA}
\affiliation{Research and Education Center for Natural Sciences, 
Keio University, Yokohama 223-8521, Japan}
\affiliation{RIKEN iTHEMS, RIKEN, Wako 351-0198, Japan}

\date{\today}

\begin{abstract}
We investigate 't Hooft anomalies in the $\CP^{N-1}$ model in spacetime dimensions higher than two and identify two types of anomalies:
One is a mixed anomaly between the $\PSU(N)$ flavor-rotation and magnetic symmetries, 
and the other is between the reflection and magnetic symmetries.
The latter indicates that even in the absence of the flavor symmetry, the model cannot have a unique gapped ground state as long as the reflection and magnetic symmetries are respected.
We also clarified the condition for the 't Hooft anomalies to survive under monopole deformations, which explicitly break the magnetic symmetry down to its discrete subgroup.
Besides, we explicitly show how the identified 't Hooft anomalies match in the low-energy effective description of symmetry broken phases---the N\'eel, $\U(1)$ spin liquid, and the valence bond solid phases.
An application to the finite-temperature phase diagram of the four-dimensional $\CP^{N-1}$ model is also discussed.
\end{abstract}

\maketitle

\tableofcontents

%%%%%%%%%%%%%%%%%%%%%%%%%
%%%%		Introduction
%%%%%%%%%%%%%%%%%%%%%%%%%
\section{Introduction} \label{sec: intro}
Symmetry provides fundamental tools to understand the non-perturbative aspects of quantum many-body systems.
An 't Hooft anomaly, an obstruction to gauge global symmetry, is such a symmetry-based theoretical approach beyond perturbative analysis.
If the quantum system under consideration has an 't Hooft anomaly, it constrains possible low-energy dynamics due to the 't Hooft anomaly matching~\cite{tHooft:1980, Frishman:1980dq, Coleman:1982yg}.
Recently, new 't Hooft anomalies involving generalized global symmetries such as discrete symmetries~\cite{Kapustin:2014lwa} and higher-form symmetries~\cite{Gaiotto:2014kfa} are found and applied to a large variety of systems both in condensed matter physics~\cite{Wen:2013oza, Kapustin:2014tfa,Cho:2014jfa,Kapustin:2014gma, Kapustin:2014lwa, Wang:2014pma, Furuya:2015coa, Tachikawa:2016cha, Thorngren:2018wwt, Jian:2017skd, Cho:2017fgz, Komargodski:2017smk, Metlitski:2017fmd,  Sulejmanpasic:2018upi, Tanizaki:2018xto, Komargodski:2017dmc, Yao:2018kel,Wen:2018zux} and high-energy physics~\cite{Gaiotto:2017yup,Tanizaki:2017bam,Shimizu:2017asf,Yamazaki:2017dra,Tanizaki:2017mtm, Guo:2017xex,Cordova:2017kue,Tanizaki:2018wtg,Wan:2018zql,Yonekura:2019vyz,Cordova:2019jnf,Cordova:2019uob,Cordova:2019bsd,Hason:2019akw,Cordova:2019wpi,Cordova:2019jqi}.
In these applications, the global anomaly induced by a large gauge transformation and a discrete symmetry transformation often plays a central role rather than the perturbative anomaly (such as the chiral anomaly) induced by an infinitesimal transformation 
(see Refs.~\cite{Witten:1982fp,Witten:1985xe} for the classic examples of the global anomaly
and also Refs.~\cite{Witten:2015aba,Witten:2019bou} for a review on a fermionic global anomaly).
Furthermore, recent developments also reveal the relation between 't Hooft anomalies and boundary physics of novel states of matter known as the symmetry-protected topological (SPT) phases~\cite{Wen:2013oza,Cho:2014jfa,Kapustin:2014lwa,Wang:2014pma}.

Quantum field theoretical approach to low-dimensional spin systems provides one representative ground where 't Hooft anomalies play a pivotal role in understanding their possible low-energy  behaviors such as properties of their ground states and energy spectra~\cite{Furuya:2015coa, Metlitski:2017fmd,  Sulejmanpasic:2018upi, Tanizaki:2018xto, Komargodski:2017smk, Komargodski:2017dmc, Yao:2018kel}. 
One can see this because the 't Hooft anomaly can be regarded as an avatar of the Lieb-Shultz-Mattis (LSM) theorem for the lattice model~\cite{Lieb:1961fr,Affleck:1986pq,Oshikawa:2000,Hastings:2003zx} 
in the continuum field theory~\cite{Cho:2017fgz, Metlitski:2017fmd}
(see also Refs.~\cite{Po:2017, Shiozaki:2018yyj, Else:2019lft} for recent developments on the LSM theorem).
%(See also Refs.~\cite{Xie:2010, Roy:2012, Parameswaran:2013, Zaletel:2014epa, Watanabe:2015, Cheng:2016pjt, Hsieh:2016emq, Song:2017, Jian:2017skd, Po:2017, Lu:2017ego, Yang:2018kua, Huang:2017, Cheng:2019uoh, Kobayashi:2018yuk, Bultinck:2018uil, Shiozaki:2018yyj,  Song:2018, Yao:2019ggu, Else:2019lft, Furuya:2019lqr} for recent developments on the LSM theorem).
Indeed, the $(1+1)$-dimensional $\CP^{1}$ model with the $\theta$ term is shown to have an 't Hooft anomaly between the $\PSU(2) (\equiv \SU(2)/\Z_{2})$ flavor symmetry (or the $\SO(3)$ spin-rotation symmetry in the condensed-matter terminology) and the charge conjugation symmetry at $\theta = \pi$~\cite{Komargodski:2017dmc}, which corresponds to the original LSM theorem for the $(1+1)$d half-integer spin chain~\cite{Metlitski:2017fmd}.
Furthermore, the $(2+1)$d $\CP^1$ model also contains the 't Hooft anomaly between the $\PSU(2)$ flavor symmetry and the $\U(1)$ magnetic symmetry (or its discrete subgroup such as the $\Z_4$ magnetic symmetry)~\cite{Komargodski:2017dmc, Komargodski:2017smk, Metlitski:2017fmd}.
This mixed 't Hooft anomaly accounts for competition between the N\'eel and valence bond solid (VBS) phases with an unconventional quantum critical point known as the deconfined quantum critical point in $(2+1)$d quantum antiferromagnets~\cite{Senthil:2004-1, Senthil:2004-2, Senthil:2005}%
\footnote{
The $\CP^{1}$ model has been also attracting attentions in the context of $(2+1)$-dimensional dualities~\cite{Wang:2017txt} (see e.g., Ref.~\cite{Senthil:2018cru} for a review).
}.

While the above anomalies ensure that the $(1+1)$d ($(2+1)$d) $\CP^1$ model shows nontrivial low-energy spectra as long as the $\PSU(2)$ flavor symmetry and charge conjugation symmetry (magnetic symmetry) are respected, it is interesting to ask whether breaking the flavor symmetry can allow the model to have a unique gapped ground state.
If one finds another anomaly without the flavor symmetry,
the anomaly ensures its ground state is still nontrivial.
In Ref.~\cite{Metlitski:2017fmd}, such 't Hooft anomalies in $(1+1)$ and $(2+1)$ dimensions are studied from the bulk SPT perspective.
Constructing bulk SPT actions, they found anomalies involving only discrete internal symmetries in the $\CP^1$ model
which are identified as lattice symmetries in the underlying lattice model (e.g., the translation and site-centered rotation symmetries).
On the other hand, Ref.~\cite{Sulejmanpasic:2018upi} directly computes an anomaly in the $(1+1)$d $\CP^{N-1}$ model
by putting the model on a nonorientable manifold and reveals that the $(1+1)$d $\CP^{N-1}$ model has an 't Hooft anomaly involving a spacetime symmetry: the one between the reflection and charge conjugation symmetries.
However, since the latter work discusses only the $(1+1)$-dimensional case, it is worthwhile to generalize their discussion to higher dimensions and explore a possible 't Hooft anomaly with a spacetime symmetry.

In this paper, we investigate 't Hooft anomalies and their matching in the $\CP^{N-1}$ model (or $N$-flavor abelian Higgs model) living on $D$-dimensional spacetime with $D\geq 3$, which captures the long-wavelength behaviors of $\SU(N)$ spin systems.
We start our discussion with an 't Hooft anomaly between the $\PSU(N) (\equiv \SU(N)/\Z_N)$ flavor symmetry and the magnetic symmetry.
Then, putting the model on a nonorientable manifold in a similar manner to Ref.~\cite{Sulejmanpasic:2018upi}, we obtain an 't Hooft anomaly involving the reflection symmetry, i.e.,
a mixed 't Hooft anomaly between the reflection symmetry and the magnetic $\U(1)$ symmetry.
Thus, the anomaly matching argument forbids the $\CP^{N-1}$ model to possess a unique gapped ground state as long as the reflection and magnetic symmetries are respected.
We also clarify the condition that the 't Hooft anomalies survive
under monopole deformations to the $\CP^{N-1}$ model, which explicitly break the magnetic $\U(1)$ symmetry down to its discrete subgroup.

We further discuss several consequences of the anomaly matching
based on the identified 't Hooft anomalies.
In the N\'eel phase, where the $\PSU(N)$ flavor symmetry is spontaneously broken,
the anomalies tell us that a topologically conserved current in this phase, or the so-called Skyrmion current, has a fractional part in the presence of background gauge fields%
\footnote{
A similar phenomenon is discussed in the context of massless QCD, where the discrete chiral symmetry plays a pivotal role.
The anomaly matching argument in the chiral symmetry breaking phase 
requires the existence of the topologically conserved current carrying the baryon number~\cite{Tanizaki:2018wtg}. 
}.
As the second application, we consider the $\U(1)$ spin liquid phase, where the magnetic $\U(1)$ symmetry is spontaneously broken, and find a topological defect carries nontrivial charges under the $\PSU(N)$ flavor and reflection symmetries.
Taking account of the possible charge-$n$ monopoles, we also discuss the case where the magnetic symmetry is explicitly broken down to its discrete subgroup $\Z_n$.
The condensation of monopoles, in this case, leads to the valence bond solid (VBS) phase, where the discrete magnetic symmetry is broken spontaneously.
The spontaneously broken discrete symmetry leads to the $n$-fold degeneracy of the ground state, 
and thus, the $(2+1)$d VBS phase can possess domain walls interpolating the degenerate ground states.
In addition to the flavor anomaly matching discussed in Ref.~\cite{Komargodski:2017smk}, 
we show that a domain wall in the $(2+1)$d VBS phase shares the same anomalies as the $(1+1)$d $\CP^{N-1}$ model via the anomaly inflow mechanism~\cite{Callan:1984sa}.
Finally, we discuss the finite-temperature phase diagram of the $\CP^{N-1}$ nonlinear sigma model in $(3+1)$ dimension.
After showing persistence of the 't Hooft anomalies in 
the finite-temperature $(3+1)$d $\CP^{N-1}$ model,
we discuss how they constrain its phase diagram
(see Ref.~\cite{Gaiotto:2017yup} for an analogous restriction in the pure Yang-Mills theory at $\theta = \pi$ and Refs.~\cite{Shimizu:2017asf, Komargodski:2017dmc, Tanizaki:2017mtm, Yonekura:2019vyz} for the discussion on other gauge theories).
We also mention its consistency with the large-$N$ analysis on the $\CP^{N-1}$ nonlinear sigma model.

This paper is organized as follows.
In Sec.~\ref{sec: global}, we briefly summarize our setup, global symmetries of the $\CP^{N-1}$ model, and its relation to quantum antiferromagnets.
In Sec.~\ref{sec: thooft}, after reviewing the 't Hooft anomaly matching argument,
we elucidate the 't Hooft anomalies in the $\CP^{N-1}$ model in general spacetime dimensions higher than two with and without the flavor symmetry.
We also investigate their stability to the monopole deformations.
In Sec.~\ref{sec: matching}, we demonstrate how the 't Hooft anomalies match in the low-energy effective theories of the possible symmetry broken phases.
Sec.~\ref{sec: 4d anom} is devoted to the discussion on
the fate of the anomalies at finite temperature and its consequence for the phase diagram of the $(3+1)$d $\CP^{N-1}$ model.
In the last section \ref{sec: summary}, we summarize our result together with the discussion on a possible realization of the mixed anomaly between the reflection symmetry and the magnetic symmetry in the lattice model.

\section{$\CP^{N-1}$ model and global symmetry}~\label{sec: global}
In this section, we first explain our setup and the global symmetries of the $\CP^{N-1}$ model (or $N$-flavor abelian Higgs model) living on a $D$-dimensional spacetime manifold $\mM_D$.
Throughout this paper, we focus on spacetime dimensions higher than two (i.e., $D>2$)
and work in Euclidean signature.

Let us introduce the $\CP^{N-1}$ model.
The Lagrangian for the $\CP^{N-1}$ model takes the following form:
\begin{align}
 \label{eq: cpn model}
 \mL_{\CP^{N-1}} = |D_a z|^2 +V(|z|^2)+\frac{1}{2 g^2} \diff a \wedge \star \diff a,
\end{align}
where $z$ and $z^\+$ denote an $N$-component complex scalar field and its Hermitian conjugate, respectively.
Notice that $|z|^2 \equiv z^\+ z$ is not constant in our model%
\footnote{
In that sense, this model is different from the $\CP^{N-1}$ \textit{nonlinear} sigma model and the one often called the $\SU(N)$ abelian Higgs model giving a linear realization of the nonlinear sigma model.
Nevertheless, only the symmetry property plays a fundamental role in our discussion, and both of the linear and nonlinear models share all the results on the 't Hooft anomalies. 
}.
We also introduced the dynamical $\U(1)$ gauge field $a \equiv a_\mu \diff x^\mu$ and the covariant derivative $D_a z = (\diff - \rmi a)z$.
The Lagrangian is invariant under the following $\U(1)$ gauge transformation:
\begin{equation}
 \begin{split}
  z(x) &\ra \rme^{\rmi \theta(x)} z(x),
  \\
  a(x) &\ra a(x) + \diff \theta(x),
 \end{split}
 \label{eq: u1 gauge}
\end{equation}
with a gauge transformation parameter $\theta(x)$.
The field strength for $a$ is given by $\diff a$, and the last term in Eq.~\eqref{eq: cpn model} represents the ordinary Maxwell term.

We then describe the global symmetries of the $\CP^{N-1}$ model.
The three symmetries, $\PSU(N)_F$, $\U(1)^{[D-3]}_M$, and $\mR$, 
which we respectively call the flavor, magnetic, and reflection symmetries, play essential roles in our discussion.
First of all, $\PSU(N)_F$ is a continuous flavor symmetry.
The $\PSU(N)$ group represents the quotient group $\SU(N)/\Z_N$, which is obtained by dividing the $\SU(N)$ group by its center:
\begin{align}
 \Z_N = \{ \rme^{2 \pi \rmi k/N } \I_N | k = 0, \cdots, N-1 \} \subset \SU(N),
\end{align}
 where $\I_N$ is the $N \times N$ identity matrix.
The flavor symmetry acts on the fields as
\begin{equation}
 \begin{cases}
  z(x) \ra U z(x),
  \\
  a(x) \ra a(x),
 \end{cases}
\end{equation}
 where $U \in \SU(N)$.
At first glance, one may na\"ively think the global symmetry group is $\SU(N)$.
However, because of the $\U(1)$ gauge invariance~\eqref{eq: u1 gauge}, elements in the center group don't act on any physical operators, and the faithful flavor symmetry is $\PSU(N)_F$ rather than $\SU(N)_F$.

The model also enjoys another continuous symmetry called the $(D-3)$-form $\U(1)$ magnetic symmetry denoted as $\U(1)^{[D-3]}_M$.
The conserved current is given by 
\begin{equation}
J_M = \frac{1}{2 \pi}\star \diff a ,
  \label{eq:J-mag}
\end{equation}
and its conservation is ensured by the Bianchi identity for the gauge field~\cite{Gaiotto:2014kfa}.
This symmetry is sometimes called a topological symmetry in the literature because the current is automatically conserved without using the equation of motion.
The generator is defined on the two-dimensional submanifold $\Sigma_2$ as
 \begin{align}
 Q_M (\Sigma_2) = \int_{\Sigma_2} \star J_M  = \frac{1}{2 \pi} \int_{\Sigma_2} \diff a.
 \end{align}
This quantity is nothing but the magnetic flux penetrating the surface $\Sigma_2$ and acts on $(D-3)$-dimensional magnetically charged objects such as monopole instantons in three dimensions and 't Hooft lines in four dimensions.

Besides, we shall also consider a situation where the magnetic symmetry is broken to its discrete subgroup $\Z_n$ because of the presence of a magnetic object with a magnetic charge $n$ in Sec.~\ref{sec: monopole}.
In three dimensions, this happens when we perform the path integral over configurations with $n$-multiple monopoles.
On the other hand, in four dimensions, we can introduce magnetic monopoles in the Maxwell theory using a $2$-form $\U(1)$ gauge field $b = \frac{1}{2}b_{\mu\nu} \diff x^\mu \wedge \diff x^\nu$, whose Lagrangian reads
\begin{align}
 \mL_{\mathrm{EM}+\mathrm{mon}} = 
 \frac{1}{2 g^2} (\diff a - n b) \wedge \star (\diff a - n b)
 + \frac{1}{8 \pi^2 \rho^2_M} \diff b \wedge \star \diff b,
\end{align}
where $\rho_M$ denotes the stiffness of the monopole condensate.
This theory represents a gauge theory coupled to charge-$n$ magnetic monopoles
because it is equivalent to the charge-$n$ scalar $\eta$ coupled to the dual vector potential $\tilde{a}$ via the S-duality 
(see 
the discussion in Sec.~\ref{sec: vbs}):
\begin{align}
 \mL_{\mathrm{EM}+\mathrm{mon}} 
 \leftrightarrow
 \frac{\rho^2_M}{2} (\diff \eta - n \tilde{a}) \wedge 
 \star (\diff \eta - n \tilde{a})
 + \frac{g^2}{ 8 \pi^2} \diff \tilde{a} \wedge \star \diff \tilde{a}.
\end{align}
We thus can interpret the scalar field $\eta$ dual to the $2$-form gauge field $b$ represents the monopole field.
The $\Z_n$ shift symmetry for $\eta$ and $\tilde{a}$ is translated 
into the magnetic symmetry in the original model defined by $\mL_{\mathrm{EM}+\mathrm{mon}} $.
In the following, we will consider the $\U(1)_M^{[D-3]}$ symmetry or the $(\Z_n)_M^{[D-3]}$ symmetry depending on whether we consider monopole deformations or not.

Third, we shall explain the reflection symmetries $\mR_{\mu}$ ($ \mu =1, \cdots, D$), which act on the fields as follows:
\begin{equation}~\label{eq: reflection}
 \begin{cases}
  z(x) \ra \Omega z^\ast(R_{\mu} x),
  \\
  a(x) \ra - (R_{\mu} \cdot a) (R_{\mu} x).
 \end{cases}
\end{equation}
Here, we introduced the $D \times D$ matrices $R_{\mu}={\rm diag}(1,\cdots,1, \overbrace{-1}^\mu, 1, \cdots, 1)$ and a certain element $\Omega \in \SU(N)$.
Note that $\mR_{1}$ represents the time-reversal transformation. 

Using the fact that $\mR_\mu$ generates a $\Z_2$ transformation, 
one can show that $\Omega$ and its transpose $\Omega^t$ satisfy (see Ref.~\cite{Sulejmanpasic:2018upi} and Appendix.~\ref{sec:Omega} for a derivation)
\begin{equation}
 \label{eq:Omega}
 \begin{split}
  &\Omega = +\Omega^t \hspace{53pt} \mathrm{for~odd~} N, \\
  &\Omega = +\Omega^t~\mathrm{or} - \Omega^t \quad \mathrm{for~even~} N.
 \end{split}
\end{equation}
For example, $\Omega = - \Omega^t$ for even $N$ can be explicitly realized by the following choice:
\begin{align}~\label{eq: omega}
 \Omega = {\rm diag} (\overbrace{\rmi \sigma^y, \cdots, \rmi \sigma^y}^{N/2} ),
\end{align}
where we used the Pauli matrix $\sigma^\alpha ~ (\alpha =x,y,z)$.
The sign in front of $\Omega^t$ in Eq.~\eqref{eq:Omega} is essential in our following discussion because it determines the presence and absence of the mixed anomaly between the reflection and magnetic symmetries.
Since these reflection symmetries are related to each other through Lorentz transformations, 
we will refer to the reflection symmetry just as $\mR$ in the next section.

Before closing this section, we comment on a relationship between the $\CP^{N-1}$ model \eqref{eq: cpn model} and spin systems in condensed matter physics.
When $N = 2$, the $\CP^1$ model enjoys the $\PSU(2)$ flavor symmetry, which can be identified as the $SO(3)$ spin-rotation symmetry of antiferromagnets because of $\PSU(2) \simeq SO(3)$.
The model is thus reduced to the $O(3)$ sigma model describing quantum antiferromagnets in the N\'eel phase after imposing the constraint $|z|^2 = \rho^2$ and 
neglecting the Maxwell term.
Note that the gauge-invariant combinations, $z^\+ \sigma^\alpha z$ with the Pauli matrices $\sigma^\alpha~(\alpha=x,y,z)$, are identified as the normalized N\'eel order parameter $n^\alpha = \rho^{-1} z^\dag \sigma^\alpha z$ with $n^\alpha n_\alpha = 1$.

On the other hand, a lattice model counterpart of the magnetic symmetry is more indirect.
In $(2+1)$ dimension, the magnetic symmetry is regarded as the spatial rotation symmetry in the lattice models, which is sensitive to their lattice structures.
For instance, the rectangular, honeycomb, and square lattices possess the $\Z_2$, $\Z_3$, and $\Z_4$ rotation symmetries, respectively, and they are identified as discrete magnetic symmetries in the presence of charge-$n$ magnetic monopoles.
This identification means that magnetic symmetry-broken phases in the $\CP^1$ model with the monopole deformations describe the VBS phases in lattice models ~\cite{Read:1989zz, Read:1990zza, Senthil:2004-2}.
Therefore, for applications to condensed matter physics, 
it is meaningful to consider the possible appearance of monopole having magnetic charge $n$, which breaks the $\U(1)^{[D-3]}_M$ magnetic symmetry down to the $(\Z_n)^{[D-3]}_M$ symmetry.

\begin{figure}[htb]
 \centering
 \includegraphics[width=0.8\linewidth]{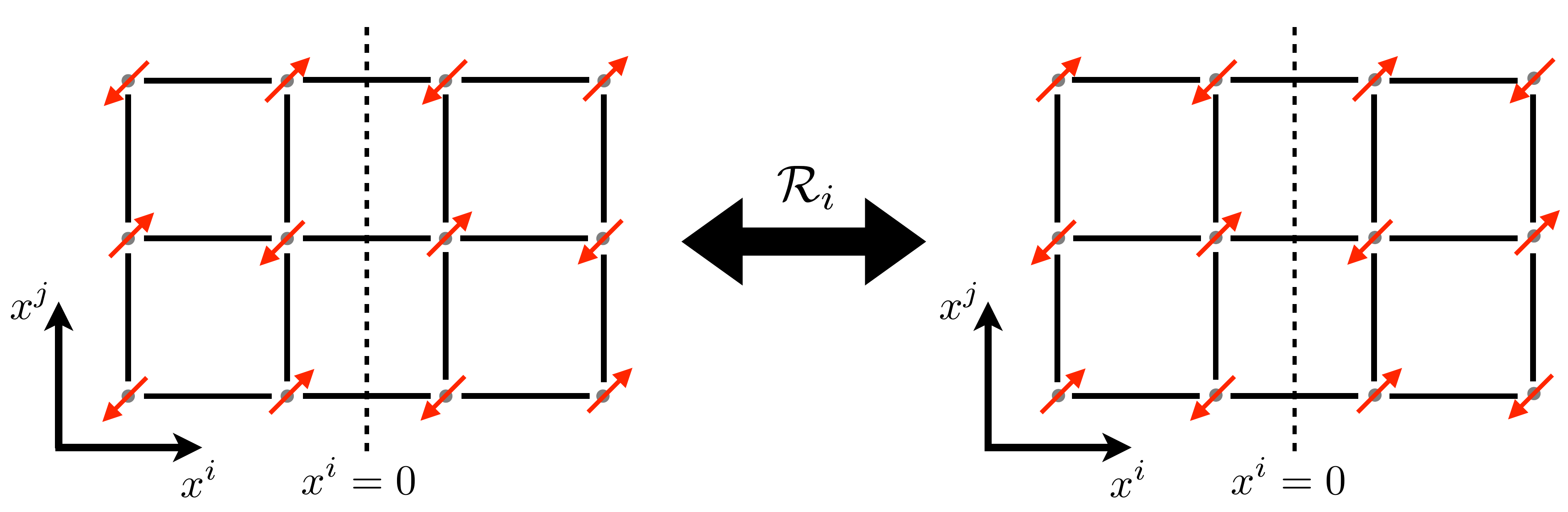}
 \caption{
 The reflection $\mR_i$ as the link-centered parity transformation along the $x^i$ direction. 
 }
\label{fig:reflection}
\end{figure}

Let us finally identify the reflections $\mR_{ \mu }$ in the antiferromagnet on, e.g., the square lattice, setting the spacetime dimension $D=3$.
In the $N=2$ flavor case, one can simply use $\Omega = \rmi \sigma^y$ satisfying $\Omega = - \Omega^t$, and the reflection $\mR_\mu$ acts on the N\'eel order parameter as
\begin{align}
 \mR_\mu:~n^\alpha (x) \ra - n^\alpha (R_\mu x).
\end{align}
This equation indicates that $\mR_1$ represents the time-reversal, and $\mR_{2}$ ($\mR_3$) does the link-centered parity in the $x^2$ ($x^3$) direction at the lattice scale (see Fig. \ref{fig:reflection}).

\section{'t Hooft anomaly}~\label{sec: thooft}
In this section, we show that the $\CP^{N-1}$ model \eqref{eq: cpn model} exhibits two 't Hooft anomalies: the one between the $\PSU(N)$ flavor and magnetic symmetries, and the other between the magnetic and reflection symmetries.
Before proceeding to it, we shall quickly review an 't Hooft anomaly and its consequence, the anomaly matching~\cite{tHooft:1980, Frishman:1980dq, Coleman:1982yg}.

Suppose that our system living on the spacetime manifold $\mM_D$ has a global symmetry $G$, such as the flavor, magnetic, and reflection symmetries in the $\CP^{N-1}$ model.
One can generally couple a background gauge field $A$ associated with the symmetry $G$ and define the generating functional for the system $Z[A]$ as
\begin{equation}
 Z[A] = \int \mD \varphi\, 
  \exp ( - S_{\mathrm{gauged}} [\varphi;A] ) ,
\end{equation}
where $\varphi$ is a set of dynamical fields---e.g., $\varphi = \{z, a\}$ in the $\CP^{N-1}$ model---and $S_{\mathrm{gauged}} [\varphi;A]$ denotes the action equipped with the background $G$-gauge field $A$.
The background gauge field promotes the global symmetry to a local one.
We na\"ively expect that the generating functional is invariant under the local transformation, $A(x) \ra A(x) + \delta_\theta A(x)$, where $\delta_\theta A(x)$ represents the gauge variation with a possible gauge parameter $\theta(x)$.
However, contrary to our na\"ive expectation, this is not always the case.
The generating functional sometimes obtains a phase factor:
\begin{align}
 Z[A + \delta_\theta A] = Z[A] \rme^{\rmi {\mathcal A}[\theta, A]},
\end{align}
and the gauge invariance is broken up to this phase.
${\mathcal A}[\theta, A]$ is a local functional of $\theta(x)$ and $A(x)$.
The system is defined to have an 't Hooft anomaly when the anomalous phase shift ${\mathcal A}[\theta, A]$ cannot be canceled out by a variation of any gauge-invariant local counterterm in $D$ dimensions (i.e., ${\mathcal A}[\theta, A] \neq \delta_\theta S^D_{\rm local}[A]$).
Instead, it is usually saturated by a variation of a local action living on $\mN_{D+1}$, a $(D+1)$d open manifold whose boundary is $\mM_{D}$:
\begin{align}
\delta_\theta S^{D+1}_{\rm SPT}[A] = \rmi {\mathcal A}[\theta, A].
\end{align}
Here $S^{D+1}_{\rm SPT}[A]$ represents the local action and describes an SPT phase protected by the symmetry $G$ on the $(D+1)$-dimensional spacetime.
We emphasize that 
the combination $Z[A] \rme^{- S^{D+1}_{\rm SPT}[A]}$ is gauge invariant
thanks to the SPT action.

If we find the 't Hooft anomaly, it inevitably constrains the low-energy behaviors of the model.
Consider the renormalization group (RG) transformation of the gauge-invariant combination $Z[A] \rme^{- S^{D+1}_{\rm SPT}[A]}$.
While the RG on the $D$-dimensional boundary 
results in a certain low-energy effective theory of the edge system,
the RG in the $(D+1)$d bulk is trivial because the SPT action has no dynamical degrees of freedom.
Thus, the low-energy effective action on the boundary must reproduce the same anomalous phase factor canceling the variation in the bulk because the RG preserves the gauge invariance of the total system.

As a result, since the effective action for any non-degenerate (unique) gapped ground state cannot reproduce the phase shift $\rmi {\mathcal A}[\theta, A]$, it is impossible to gap out the system without ground state degeneracy.
Therefore, possible scenarios for the low-energy behaviors of systems (such as ground state) are restricted to show
\begin{itemize}
\item \textit{spontaneous symmetry breaking of $G$},
\item \textit{topological order}, or
\item \textit{conformal behavior},
\end{itemize}
where a unique gapped ground state is ruled out.
This consistency condition on the infrared (IR) behaviors of the system is known as the 't Hooft anomaly matching.

In order to detect the 't Hooft anomaly, we thus need to gauge global symmetries in the $\CP^{N-1}$ model: the flavor, magnetic, and reflection symmetries.
Since gauging the flavor and reflection symmetries requires some effort, we first explain a straightforward part, gauging the $\U(1)_M^{[D-3]}$ magnetic symmetry%
\footnote{
See Sec.~\ref{sec: monopole} for an extension to the case with charge-$n$ monopoles, where the magnetic symmetry is broken to be $(\Z_n)_M^{[D-3]}$.}. 
It is accomplished by introducing a background $(D-2)$-form gauge field $K$ through the minimal coupling to the magnetic current $J_M$ defined in Eq.~\eqref{eq:J-mag}.
Then, the action for the $\CP^{N-1}$ model takes the following form:
\begin{align}
 \label{eq:Gauged-mag}
 S [z,a;K] = 
 \int \left[ |D_a z|^2 +V(|z|^2) +\frac{1}{2 g^2} \diff a \wedge \star \diff a \right]
 +\frac{\rmi}{2 \pi} \int K\wedge \diff a,
\end{align}
where the last term results from gauging the $\U(1)_M^{[D-3]}$ symmetry.
As a result, the action \eqref{eq:Gauged-mag} becomes invariant under the $\U(1)^{[D-3]}_M$ local transformation given by
\begin{equation}
\label{eq: u1m}
 \begin{cases}
  z(x) \ra z(x),
  \\
  a(x) \ra a(x),
  \\
  K(x) \ra K(x) + \diff \theta(x),
 \end{cases}
\end{equation}
with a $(D-3)$-form local parameter $\theta(x)$.
Under this transformation, the generating functional is invariant because the action changes only by
\begin{align}~\label{eq: u1m inv}
\frac{\rmi}{2 \pi} \int \diff \theta \wedge \diff a \in 2 \pi \rmi \Z, 
\end{align}
which doesn't affect the generating functional.
Note that this gauge field $K$ doesn't transform under the flavor symmetry but changes under $\mR_\mu$ as 
\begin{align}~\label{eq: cp trsf k}
 K(x) \ra  (R_{\mu} \cdot K)(R_{\mu}x).
\end{align}
We demand this transformation to keep $\frac{\rmi}{2 \pi} \int K\wedge \diff a$ invariant under the reflection $\mR_\mu$.

At this stage, we do not encounter any obstruction to gauge the magnetic symmetry, but the situation will be changed when we try to gauge other symmetries.
In the following, we study the $\PSU(N)_F \times \U(1)^{[D-3]}_M$ and $\mR \times \U(1)^{[D-3]}_M$ anomalies in Sec.~\ref{sec: psu anom} and Sec.~\ref{sec: cp anom}, respectively.
We will observe that turning on background gauge fields for $\PSU(N)_F$ and $\mR$ spoils the $\U(1)^{[D-3]}_M$ invariance~\eqref{eq: u1m inv}, which implies the existence of the mixed 't Hooft anomalies.
In Sec.~\ref{sec: monopole}, we also discuss whether the 't Hooft anomalies exist or not when $\U(1)^{[D-3]}_M$ is broken down to its discrete subgroup by the monopole deformation of the model.

\subsection{$\PSU(N)_F \times \U(1)^{[D-3]}_M$ anomaly}~\label{sec: psu anom}

Let us begin with coupling the system to a background gauge field for the $\PSU(N)_F$ symmetry~\cite{Kapustin:2014gua, Gaiotto:2014kfa}.
The $\PSU(N)$ gauge field is realized as a pair of a $\U(N)$ gauge field $A$ and a $2$-form $\U(1)$ gauge field $B$ satisfying the following constraint: 
\begin{align} \label{eq: constraint}
N B = \tr F[A],
\end{align}
where $F[A] = \diff A - \rmi A \wedge A$ is the field strength for $A$.
This constraint allows us to regard $B$ as a $2$-form $\Z_N$ gauge field,
whose surface integration is fractionally quantized:
\begin{align}~\label{eq: b quant}
\int_{\Sigma_2} B \in \frac{2 \pi}{N} \Z,
\end{align}
where an arbitrary two-dimensional closed manifold $\Sigma_2$.
The $\U(N)$ gauge field acts on the fields in the ordinary manner:
\begin{equation}~\label{eq: u1 1form}
 \begin{cases}
  z(x) \ra U(x) z(x),
  \\
  a(x) \ra a(x),
  \\
  K(x) \ra K(x),
  \\
  A(x) \ra U(x)A(x) U^\+(x) - \rmi \diff U(x) U^\+(x) ,
  \\
  B(x) \ra B(x),
 \end{cases}
\end{equation}
with a gauge transformation parameter $U(x) \in \U(N)$.
Nevertheless, since the $\U(N)$ gauge field contains the undesired $\U(1)$ component, we eliminate this by imposing the $1$-form gauge invariance:
\begin{equation}~\label{eq: u1 1form}
 \begin{cases}
  z(x) \ra z(x),
  \\
  a(x) \ra a(x) - \lambda(x),
  \\
  K(x) \ra K(x),
  \\
  A(x) \ra A(x) + \lambda(x) \I_N ,
  \\
  B(x) \ra B(x) + \diff \lambda (x),
 \end{cases}
\end{equation}
where $\lambda(x)$ is a $1$-form $\U(1)$ field.
Note that one needs to transform the $2$-form gauge field $B$
to keep the constraint \eqref{eq: constraint}.
This $1$-form gauge invariance removes the redundant diagonal part in $\U(N)$, which acts on no gauge-invariant operators as discussed in Sec.~\ref{sec: global}, and allows us to realize the $\U(N)/\U(1) \simeq \PSU(N)$ gauge field.
It may be also helpful to recall a relation $\U(N) \simeq \U(1) \times \SU(N) /\Z_N$, 
where the center of $\SU(N)$ is divided to avoid double counting with the $\U(1)$ part.

After introducing the $\PSU(N)_F$ gauge field, the resulting action of the $\CP^{N-1}$ model takes the form:
\begin{equation}
 \begin{split}
  S [z,a;A,B,K]
  = \int \Bigl[ |D_{a+A} z|^2 +V(|z|^2) 
  + \frac{1}{2 g^2} (\diff a+B) \wedge \star (\diff a+B) 
  + \frac{\rmi}{2 \pi}  K\wedge (\diff a+B) \Bigr].
 \end{split}
\end{equation}
The covariant derivative for $z$ is now replaced by $D_{a+A}z = (\diff - \rmi a - \rmi A)z$.
This completes gauging the quotient group symmetry $\PSU(N)_F$.

We then show gauging $\PSU(N)_F$ indeed spoils the large gauge invariance for $\U(1)^{[D-3]}_M$.
The generating functional similarly acquires an anomalous phase factor under the local $\U(1)^{[D-3]}_M$ transformation, $K(x) \ra K(x) + \diff \theta(x)$ as
\begin{align}~\label{eq: psu anom Ddim}
Z[K + \diff \theta ,A,B] = Z[K,A,B] 
 \exp \left( - \frac{ \rmi }{2 \pi} \int_{\mM_D} \diff \theta \wedge B \right).
\end{align}
Thus, a nontrivial phase shift under the large gauge transformation remains 
because the fractional quantization~\eqref{eq: b quant} yields $\frac{ \rmi }{2 \pi} \int_{\mM_D} \diff \theta \wedge B \in \frac{2\pi \rmi}{N} \Z$.

We must confirm no local counterterm cancels the phase shift.
A possible local action canceling the variation takes the form:
\begin{align}
\int_{\mM_D} K \wedge B.
\end{align}
Nevertheless, it is not invariant under the $\Z_N$ $1$-form gauge transformation,
and we have no local counterterm to cancel the phase shift.
Therefore we find the $\PSU(N)_F \times \U(1)^{[D-3]}_M$ anomaly
in the $D$-dimensional $\CP^{N-1}$ model.
Furthermore, one can show this anomaly is saturated by attaching the $\CP^{N-1}$ model on the boundary of an SPT phase living on a $(D+1)$-dimensional manifold $\mN_{D+1}$:
\begin{align}
 S_{\rm SPT}[K,A,B] = (-1)^{D+1}\frac{\rmi}{2 \pi} \int_{\mN_{D+1}} K \wedge \diff B.
\end{align}
One can readily show the combination $Z[K,A,B] e^{- S_{\rm SPT}[K,A,B]}$ is invariant under both Eq.~\eqref{eq: u1m} and Eq.~\eqref{eq: u1 1form}.

\subsection{$\mR \times \U(1)^{[D-3]}_M$ anomaly}~\label{sec: cp anom}
We next study the mixed anomaly between the reflection symmetry $\mR$ and the magnetic symmetry $\U(1)^{[D-3]}_M$.

To detect the anomaly, we need to introduce a gauge field for $\mR$.
A flux for such a discrete symmetry is introduced by imposing a twisted boundary condition associated with the symmetry.
In the case of the reflection symmetry, it is equivalent to putting the theory on a nonorientable manifold (see Refs.~\cite{Thorngren:2018wwt, Sulejmanpasic:2018upi} and also related work~\cite{Kapustin:2014tfa,Kapustin:2014gma}).

For simplicity, we first consider the $\CP^{N-1}$ model living on a three-dimensional cube whose side length is $L$.
We choose the periodic boundary condition along the $x^1$ direction 
but impose the twisted boundary conditions on the other ($x^2$ and $x^3$) directions by using the $\mR_{3}$ and $\mR_{2}$ transformations.
The boundary condition along the $x^2$ direction is explicitly given by
\begin{subequations} % 2019-11-9 01:21equations
 \begin{align}
  z(x^1, L/2, x^3) &= \rme^{\rmi \theta_2 (x^1,x^3)} \Omega z^\ast (x^1, - L/2, -x^3),
  \\
  a(x^1, L/2, x^3) &= - (R_{3} \cdot a) (x^1, - L/2, -x^3)+ \diff \theta_2 (x^1,x^3),
  \\
  K(x^1, L/2, x^3) &= - (R_{3} \cdot K) (x^1, - L/2, -x^3),
 \end{align}
\end{subequations}
and that along the $x^3$ directions is 
\begin{subequations} % 2019-11-9 01:26equations
 \begin{align}
  z(x^1, x^2, L/2) &= \rme^{\rmi \theta_3 (x^1,x^2)} \Omega z^\ast (x^1, -x^2, -L/2),
  \\
  a(x^1, x^2, L/2)&= - (R_{2} \cdot a) (x^1, -x^2, -L/2) + \diff \theta_3 (x^1,x^2),
  \\
  K(x^1, x^2, L/2) &= - (R_{2} \cdot K) (x^1, -x^2, -L/2),
 \end{align}
\end{subequations}
where we introduced possible  $\U(1)$ phases $\theta_2$ and $\theta_3$ allowed by the gauge invariance~\eqref{eq: u1 gauge}.
We will specify the condition imposed on these phases resulting from consistency of the twisted boundary conditions.
Note that imposing these boundary conditions is equivalent to putting the theory on the $S^1 \times \RP^2$ manifold.

Let us specify a condition which the $\U(1)$ phases $\theta_1$ and $\theta_2$ satisfy.
Recalling that we have two ways to patch the scalar field on the line $(x^1, L/2, L/2) = (x^1, - L/2, - L/2)$ as 
\begin{equation}\label{eq: constraint 1}
 \begin{split}
  z(x^1, L/2, L/2)  
  &= \rme^{\rmi \theta_2 (x^1,L/2)} \Omega z^\ast (x^1, -L/2, -L/2)
  \\
  &= \rme^{\rmi \theta_3 (x^1,L/2)} \Omega z^\ast (x^1, - L/2, -L/2),
 \end{split}
\end{equation}
we find that $\theta_2 (x^1 ,L/2)$ is equal to $\theta_3 (x^1,L/2)$ modulo $2 \pi$.
In a similar way, the consistency on the line $(x^1, L/2, - L/2) = (x^1,-L/2, L/2)$ gives us
\begin{equation} \label{eq: constraint 2}
 \begin{split}
  z(x^1, L/2, - L/2) 
  &= \rme^{\rmi \theta_2 (x^1, - L/2)} \Omega z^\ast (x^1, - L/2, L/2)
  \\
  &= \rme^{\rmi \theta_3 (x^1,-L/2)} \Omega^t z^\ast (x^1, - L/2 , L/2).
 \end{split}
\end{equation}
Eq.~\eqref{eq: constraint 2} indicates
the $\U(1)$ phases satisfy
\begin{align}
\theta_2 (x^1, - L/2) - \theta_3 (x^1, -L/2) 
 \in 
 % \left\{
 % \begin{matrix} 
 \begin{cases}
  2 \pi \Z &  \ \ \  {\rm for} \ \Omega \Omega^\ast = +1,
  \\
  \pi + 2 \pi \Z &  \ \ \ {\rm for} \ \Omega \Omega^\ast = -1,
 \end{cases}
 % \end{matrix}
 % 	\right. 
\end{align}
which plays a pivotal role in finding the $\mR \times \U(1)_M^{[D-3]}$ anomaly.

We then calculate the Dirac quantization condition for the dynamical gauge field $a$ on the $\RP^2$ plane and show the $\mR \times \U(1)_M^{[D-3]}$ anomaly.
With the help of Stokes' theorem and the twisted boundary conditions,
the magnetic flux threading the plane is directly evaluated as
\begin{align}
\int_{x^2, x^3} \diff a 
 = &- \int^{L/2}_{- L/2} \diff x^2 
 \Big[  a_2 (x^1, x^2, L/2 ) -  a_2 (x^1, -x^2, -L/2 )  \Big]
 \nonumber \\
 &+ \int^{L/2}_{- L/2} \diff x^3 
 \Big[ a_3 (x^1, L/2, x^3) -  a_3 (x^1, -L/2, -x^3) \Big]
 \nonumber
\\
 = &- \left[ \theta_2 ( x^1, - L/2)  -\theta_3 (x^1,- L/2) \right]
 \nonumber \\
 &+ \left[  \theta_2 ( x^1, L/2) - \theta_3 (x^1, L/2)  \right].
 \nonumber
\end{align}
Thus, the choice $\Omega \Omega^\ast = +1$ results in the standard Dirac quantization condition,
while $\Omega \Omega^\ast = -1$ brings about a nontrivial one:
\begin{align}
 &\int_{x^2, x^3} \diff a 
 \in 2 \pi \Z ,
 \hspace{31pt} 
 \mathrm{for} \ \ \Omega \Omega^\ast  = +1 ,
 \\
 &\int_{x^2, x^3} \diff a  \in 2 \pi \Z + \pi ,
 \ \ \ 
 \mathrm{for} \ \ \Omega \Omega^\ast  = -1 .
\end{align}
Here, the latter result indicates
\begin{align}
 \int_{x^2, x^3} \diff a = \int_{x^2, x^3} \pi w_2 \ \ \ ({\rm mod \  2 \pi}),
\end{align}
where $w_2$ denotes the second Stiefel-Whitney class for the tangent bundle on our spacetime manifold~\cite{Milnor:2016} (see also~\cite{Kapustin:2014gma, Kapustin:2014tfa}).
It is defined modulo $2$ and takes zero and one on spin and non-spin manifolds, respectively.
In that sense, it does not directly measure the orientability of the manifold, which is instead detected by the first Stiefel-Whitney class. 
However, since we introduce the twisted boundary conditions just in the two directions,
the second Stiefel-Whitney class is related to the first Stiefel-Whitney class via $w_2 = w_1^2$ (see the appendix of Ref.~\cite{Kapustin:2014gma}), and the nontrivial second Stiefel-Whitney class means nonorientablity of the manifold.
As will be discussed shortly, it is worth emphasizing that $\pi w_2$ plays a similar role to the $2$-form gauge field $B$ in Sec.~\ref{sec: psu anom}.

The modified Dirac quantization condition might imply the $\CP^{N-1}$ model has an 't Hooft anomaly.
Under the twisted boundary conditions with $\Omega \Omega^\ast= -1$,
performing the large $\U(1)_M$ gauge transformation $K(x) \ra K(x) + \diff \theta(x)$ with $\theta(x) = 2 \pi n x^1/L$ ($n \in \Z$) gives us 
\begin{align}~\label{eq: cp large gauge}
 Z[K+ \diff \theta, w_2]
 = Z[K, w_2] 
 \exp \left( - \rmi n \int_{x^2, x^3} \pi w_2 \right).
\end{align}
Thus, the choice $\Omega \Omega^\ast= -1$ could yield a mixed 't Hooft anomaly between the $\mR$ and $\U(1)_M$ symmetries while one does not find such an anomaly in the other case $\Omega \Omega^\ast= +1$.

A possible local term which cancels the anomalous phase~\eqref{eq: cp large gauge} is just
\begin{align}~\label{eq: local counter cp}
- \frac{\rmi}{2 \pi} \int_{\mM_3} K \wedge \pi w_2.
\end{align}
However, one finds that this term is ill-defined as follows.
For example, when we substitute a specific configuration $K = \eta \diff x^1/L$ with $\eta \in [0,2\pi]$ into Eq.~\eqref{eq: local counter cp} and put the theory on the manifold $S^1 \times \RP^2$, this term \eqref{eq: local counter cp} reduces to
\begin{align}
- \frac{\rmi \eta}{2 \pi} \int_{x^2,x^3} \pi w_2,
\end{align}
which is not consistent for a generic value of $\eta \in [0, 2\pi]$ because $\int_{x^2,x^3} w_2$ is only defined modulo $2$.
Therefore, we cannot remove the anomalous phase shift~\eqref{eq: cp large gauge} 
by adding any local counterterms, which proves that the $\CP^{N-1}$ model possesses the $\mR \times \U(1)_M$ anomaly.
In addition, the SPT action corresponding to this anomaly is identified as 
\begin{align}~\label{eq: spt cp}
 S_{\rm SPT}[K, w_2] = \frac{\rmi}{2 \pi} \int_{\mN_4} K \wedge \pi \diff w_2.
\end{align}
Likewise, $\mN_4$ is a four-dimensional bulk whose boundary is given by $\mM_3 = S^1 \times \RP^2$.

Generalizing the above discussion to higher dimensions is straightforward.
We impose the twisted boundary conditions on two directions,
or put the model on a manifold $\mM_D = \mM_{D-2} \times \RP^2$.
When one chooses $\Omega\Omega^\ast = -1$, the generating functional similarly picks up the following phase factor under the local $\U(1)^{[D-3]}_M$ transformation:
\begin{align}~\label{eq: cp anom Ddim}
 Z[K + \diff \theta, w_2] = Z[K, w_2] 
 \exp \left( - \frac{\rmi}{2\pi} \int \diff \theta \wedge\pi w_2 \right).
\end{align}
No local counterterms saturate this phase factor, but the following SPT action living on a $(D+1)$-dimensional manifold $\mN_{D+1}$ with $\partial \mN_{D+1} = \mM_D$ cancels it:
\begin{align}~\label{eq: spt action cp}
 S_{\rm SPT}[K, w_2] = (-1)^{D+1}\frac{\rmi}{2 \pi} \int_{\mN_{D+1}} K \wedge \pi \diff w_2.
\end{align}
Thus, we find the $\mR \times \U(1)^{[D-3]}_M$ anomaly in general dimensions.
Note that $\Omega\Omega^\ast = +1$ gives no anomaly again.

Here, we mention a relation to the previous work.
Up to a total derivative term, the SPT action~\eqref{eq: spt cp} is rewritten as
\begin{align}
 S_{\rm SPT}[K, w_2] = - \frac{\rmi}{2 \pi} \int_{\mN_4} \diff K \wedge \pi w_2.
\end{align}
This SPT action coincides with one of the topological actions for topological paramagnets protected by the $\U(1) \times \mT$ symmetry~\cite{Kapustin:2014tfa}.
Since the transformation law of $K$ given in Eq.~\eqref{eq: cp trsf k} under $\mR \sim \mR_{1} \sim \mT$
is different from that of the ordinary $\U(1)$ gauge field by a charge conjugation,
it obeys the classification of topological paramagnets.
We note that our SPT action~\eqref{eq: spt action cp} also describes its higher-form generalizations.

In summary, we find the $\mR \times \U(1)^{[D-3]}_M$ anomaly in the $\CP^{N-1}$ model  when the number of flavor $N$ is even, and we further choose $\Omega \Omega^\ast = -1$.
The $\mR \times \U(1)^{[D-3]}_M$ is obtained without using the $\PSU(N)_F$ symmetry.
Therefore, the $\mR \times \U(1)^{[D-3]}_M$ anomaly survives even if we allow the model to have any potential for the scalar field breaking the flavor symmetry but respecting the reflection symmetry.
This indicates that even when the model explicitly breaks the flavor symmetry, it cannot show a unique gapped ground state as long as the reflection and magnetic symmetries are respected.
Furthermore, we also emphasize that the $\mR \times \U(1)_M^{[D-3]}$ anomaly is present even without the Lorentz or rotation symmetries but only with the magnetic symmetry and two of the $\mR_{\mu}$ ($\mu = 1, \cdots, D$) symmetries.
% However, this is not the case when $\Omega$ satisfies $\Omega \Omega^\ast=+1$.

\subsection{Stability to monopole deformation}~\label{sec: monopole}
In this subsection, we investigate stability of the above anomalies under the monopole deformation, which explicitly breaks the magnetic symmetry.
When the model couples to a charge-$n$ magnetic object,
the magnetic symmetry is broken down to its $\Z_n$ subgroup referred to as $(\Z_n)^{[D-3]}_M$.
In this case, the background gauge field is replaced by a $(D-2)$-form $\Z_n$ gauge field $K_n$, which is realized as a $(D-2)$-form $\U(1)$ gauge field satisfying the constraint $n K_n = \diff H$ with $(D-3)$-form $\U(1)$ gauge field $H$.
Due to this constraint,
the local $(\Z_n)^{[D-3]}_M$ transformation acts on both $K_n$ and $H$ as 
\begin{equation}
 \label{eq: zn gauge trsf}
  \begin{cases}
  K_n \ra K_n + \diff \theta ,
  \\
  H \ra H + n \theta,
  \end{cases}
\end{equation}
where $\theta$ denotes a $(D-3)$-form local function.
Although the gauged action itself takes the same form up to the replacement of $K$ with $K_n$ (and an action for a magnetic object),
this deformation allows us to have additional local counterterms, which may cancel the anomalous phases, Eqs.~\eqref{eq: psu anom Ddim} and~\eqref{eq: cp anom Ddim}.
If they cancel the anomalous phase shifts, the anomalies are not stable to the explicit breaking of $\U(1)^{[D-3]}_M$, and such a deformation results in a unique gapped ground state.
In the following, we clarify conditions for the anomalies to survive by
investigating whether we can remove the anomalous phase shifts with the new local counterterms or not.

First, we begin with the $\PSU(N)_F \times (\Z_n)^{[D-3]}_M$ anomaly.
Thanks to the $(D-3)$-form gauge field $H$, we also have the local counterterm:
\begin{align}
S_k[H,B] = \frac{\rmi k}{2 \pi} \int \diff H \wedge B,
\end{align}
which possibly cancels the anomalous phase shift in Eq.~\eqref{eq: psu anom Ddim}.
Here, note that the coefficient $k$ is an integer running from $0$ to $N-1$ because $B$ denotes a $\Z_N$ gauge field.
Then, let $Z[K_n, H, A, B]$ be the generating functional for the $\CP^{N-1}$ model with the monopole deformation, which acquires the same phase factor as the original one~\eqref{eq: psu anom Ddim}
under the $(\Z_n)^{[D-3]}_M$ transformation~\eqref{eq: zn gauge trsf}.
With the help of the new local counterterm, we redefine the generating functional as
$\widetilde{Z}[K_n, H, A, B] = Z[K_n, H, A, B] \rme^{- S_k[H,B]}$, which results in
\begin{equation}
 \begin{split}
  \widetilde{Z} & [K_n+\diff \theta, H+n \theta, A, B]
  \\
  &= \widetilde{Z}[K_n, H , A, B] \exp\left(- \frac{\rmi (kn + 1 )}{2 \pi} \int \diff \theta \wedge B \right).
 \end{split}
\end{equation}
This equation means $S_k[H,B]$ can saturate the anomaly if and only if $(kn + 1) \in N\Z$.
In other words, such $S_k[H,B]$ exists when $\mathrm{GCD}(n,N) = 1$.
We thus conclude the flavor anomaly vanishes in this case, but otherwise it is still present, which reproduces the previous result in Ref.~\cite{Komargodski:2017dmc} for $D=3$.

Let us next clarify the condition under which the $\mR \times (\Z_n)^{[D-3]}_M$ anomaly survives.
A parallel discussion allows the additional local counterterm:
\begin{align}
S_k[H,w_2] = \frac{\rmi k}{2 \pi} \int \diff H \wedge \pi w_2,
\end{align}
where $k=0,1$ resulting from the property of the second Stiefel-Whitney class.
We then modify the generating functional for the deformed model $Z[K_n, H, w_2]$ by adding this local counterterm as
$\widetilde{Z}[K_n, H, w_2] = Z[K_n, H, w_2]e^{- S_k[H,w_2]}$, whose large $(\Z_n)_M^{[D-3]}$ gauge transformation results in 
\begin{equation}
 \begin{split}
  \widetilde{Z} & [K_n+\diff \theta, H+n \theta, w_2]
  \\
  &=\widetilde{Z}[K_n, H, w_2]
  \exp \left(-\frac{\rmi (kn + 1 )}{2 \pi} \int \diff \theta \wedge w_2 \right).
 \end{split}
\end{equation}
This shows the $\mR \times (\Z_n)^{[D-3]}_M$ anomaly exists for even $n$,
while it is a fake for odd $n$.

\section{Anomaly matching in low-energy effective theories}~\label{sec: matching} 
In this section, we investigate the 't Hooft anomaly matching of the $\CP^{N-1}$ model \eqref{eq: cpn model} in the low-energy effective theory showing spontaneous symmetry breaking.
In Sec.~\ref{sec: scenarios}, we first demonstrate possible scenarios with spontaneous symmetry breaking which saturates the specified 't Hooft anomalies.
Then, we heuristically show how 
the low-energy effective theories in the symmetry-breaking phases--- the N\'eel, $\U(1)$ spin liquid, and VBS phases---consistently match the specified 't Hooft anomalies in Secs.~\ref{sec: neel}, \ref{sec: u1 spin}, and \ref{sec: vbs}, respectively.

\subsection{Possible symmetry breaking scenarios}~\label{sec: scenarios}
Let us first constrain possible ground states with spontaneous symmetry breaking for the $\CP^{N-1}$ model in the absence of monopoles.
The model without monopoles enjoys the topological $\U(1)_M^{[D-3]}$ symmetry, and Table \ref{tbl: anom wo monopole} demonstrates a short summary on the 't Hooft anomalies obtained in the previous section.
Recall that the $\mR \times \U(1)_M^{[D-3]}$ anomaly is sensitive to whether the number of the flavors $N$ is even or odd and how $\mR$ is represented in contrast to the $\PSU(N)_F \times \U(1)_M^{[D-3]}$ anomaly.
%%%%%%%%%%%%%%%%%%%%%
%%%		table
%%%%%%%%%%%%%%%%%%%%%
\begin{table}[t]
\centering
\begin{tabular}{c | c  c}
\hline \hline 
\ & $\PSU(N)_F \times \U(1)^{[D-3]}_M$ & $\mR \times \U(1)^{[D-3]}_M$
\\ \hline
$\Omega = + \Omega^t$ & $\bigcirc$ & $\times$
\\ \hline
$\Omega = - \Omega^t$ & $\bigcirc$ & $\bigcirc$
\\
\hline \hline
\end{tabular}
 \caption{
 't Hooft anomalies for the $\CP^{N-1}$ model with $\U(1)^{[D-3]}_M$ ($\bigcirc$ and $\times$ indicate the presence and absence of an anomaly). } 
\label{tbl: anom wo monopole}
\end{table}

In this case, the 't Hooft anomaly matching tells us that we have the following two typical symmetry breaking scenarios:
\begin{itemize}
\item \textit{spontaneous $\PSU(N)_F$ symmetry breaking} (N\'eel phase) and
\item \textit{spontaneous $\U(1)^{[D-3]}_M$ symmetry breaking} ($\U(1)$ spin liquid phase).
\end{itemize}
We call the $\PSU(N)$ symmetry-broken phase as the N\'eel phase---indeed, the case with $N = 2$ represents the N\'eel phase for the antiferromagnet---and the $\U(1)_M^{[D-3]}$ symmetry broken phase as the $\U(1)$ spin liquid phase. 

These phases can be realized by tuning the potential term in the action~(\ref{eq: cpn model}).
For example, suppose that the potential be
$V(|z|^2) = m |z|^2 + \lambda |z|^4$
with $\lambda >0$.
Then, for $m<0$, the complex scalars can condense, and an order parameter $\Xi_{ab} = z_a z^\ast_b $ acquires non-zero expectation value.
The nonvanishing order parameter $\Xi_{ab}$ breaks the $\PSU(N)_F$ symmetry as well as the reflection symmetry.
On the other hand, for $m>0$, the complex scalars are expected to be gapped, and thus, we can integrate out them.
In this case, the dynamical gauge field $a$ remains massless, which means the spontaneous breaking of the $\U(1)^{[D-3]}_M$ symmetry.
This is why we refer to this $ \U(1)_M^{[D-3]}$ symmetry broken phase as the $\U(1)$ spin liquid phase.
\medskip

We next consider the charge-$n$ monopole deformation, which explicitly breaks
 the $\U(1)^{[D-3]}_M$ symmetry down to its discrete subgroup $(\Z_n)^{[D-3]}_M$.
Table~\ref{tbl: anom w monopole} shows the $\CP^{N-1}$ model in the presence of charge-$n$ monopoles has a little complicated anomaly structure depending on the value of the monopole charge $n$.
The N\'eel phase and $\U(1)$ spin liquid phases also saturate the $\PSU(N)_F \times  (\Z_n)_M^{[D-3]}$ and $\mR \times (\Z_n)_M^{[D-3]}$ anomalies
since $(\Z_n)_M^{[D-3]}$ is included in $\U(1)^{[D-3]}_M$.
Besides,
the model can have a spontaneous $(\Z_n)_M^{[D-3]}$ symmetry-broken phase.
We call the $(\Z_n)_M^{[D-3]}$ symmetry broken phase as a valence bond solid (VBS) phase following a terminology often used in $D=3$.
It is worth to emphasize that this phase describes \textit{topological order} in $D \ge 4$ since the broken symmetry is a higher-form discrete one~\cite{Gaiotto:2014kfa,Wen:2018zux}. 
The model allows the additional symmetry breaking scenario:
\begin{itemize}
\item \textit{spontaneous symmetry breaking of $(\Z_n)^{[D-3]}_M$} (valence bond solid (VBS) phase).
\end{itemize}
%%%%%%%%%%%%%%%%%%%%%
%%%		table
%%%%%%%%%%%%%%%%%%%%%
\begin{table}[t]
\centering
\begin{tabular}{c | c | c  c }
\hline \hline
\ & monopole charge $n$ & $\PSU(N)_F \times (\Z_n)^{[D-3]}_M$ & $\mR \times (\Z_n)^{[D-3]}_M$ 
\\ \hline
\multirow{2}{2cm}{$\Omega = + \Omega^t$} 
	& ${\rm GCD}(N, n) = 1$ &$\times$ & $\times$ 
\\ %\cline{2-4}
	& ${\rm GCD}(N, n) > 1$ &$\bigcirc$ & $\times$ 
\\ \hline
\multirow{3}{2cm}{$\Omega = - \Omega^t$}  
	& ${\rm GCD}(N, n) = 1$,  ${\rm GCD}(2, n) = 1$ & $\times$ & $\times$ 
	\\ %\cline{2-4}
	& ${\rm GCD}(N, n) > 1$,  ${\rm GCD}(2, n) = 1$ & $\bigcirc$ & $\times$ 
	\\ %\cline{2-4}
	& ${\rm GCD}(N, n) > 1$,  ${\rm GCD}(2, n) > 1$ & $\bigcirc$ & $\bigcirc$ 
\\
\hline \hline
\end{tabular}
 \caption{'t Hooft anomalies for the $\CP^{N-1}$ model in the presence of charge-$n$ monopoles ($\bigcirc$ and $\times$ indicate the presence and absence of an anomaly). 
 } 
\label{tbl: anom w monopole}
\end{table}

The 't Hooft anomaly matching states the low-energy effective action associated with the above symmetry broken phases must reproduce the anomalous phase shifts. 
Although all the above scenarios with spontaneous symmetry breaking could saturate the 't Hooft anomaly, it is quite nontrivial 
how the 't Hooft anomalies are saturated in the low-energy effective theory.
In the rest of this section, we directly examine how the saturation of the anomaly is realized in the low-energy effective action describing the symmetry-broken phases.

\subsection{N\'eel phase}~\label{sec: neel}
In this subsection, we demonstrate the anomaly matching in the low-energy effective field theory of the N\'eel phase, where $\PSU(N)_F$ is spontaneously broken to the $\U(N-1)_F$ symmetry.

As we explained in the previous subsection, the potential 
$V (|z|^2) = m |z|^2 + \lambda |z|^4 $ with $m<0$ may result in the condensation of $z_i$ in its ground state.
We then assume this condensation indeed occurs and then, parametrize the ground state expectation value as e.g., $\braket{z_a} = f_\pi \delta_a^1$ with a real constant $f_\pi$ (classically given by $f_\pi = \sqrt{- m/\lambda}$) and the Kronecker delta $\delta^a_b$.
This ground state expectation remains invariant under the act of the subgroup $\U(N-1)_F$.
Further taking account of the $\U(1)$ gauge invariance, we
see that the actual broken symmetry forms $\CP^{N-1} \simeq \U(N)/[\U(1) \times \U(N-1)]$.
Then, the resulting Nambu-Goldstone mode is introduced as 
\begin{equation}
 z (x) = U(x) \braket{z}
  \quad \mathrm{with} \quad 
  U(x) \in \U(N)/[\U(1) \times \U(N-1)],
\end{equation}
where the coset $U(x)$ contains the Nambu-Goldstone mode.
Note that we took account of the fact that $\braket{z}$ is invariant under the unbroken symmetry $\U(N-1)_F$ and chose the unitary gauge. 
Substituting this into the original Lagrangian \eqref{eq: cpn model}, we obtain 
\begin{equation}
 \mL
  = f_\pi^2 \tr 
  \left[
   \Xi^0
   \left( \diff U^\dag \wedge \star \diff U 
    + \rmi a \wedge \star( U^\dag \diff U - \diff U^\dag U) \right)
  \right]
  + \frac{1}{2g^2} \diff a \wedge \star \diff a
  + f_\pi^2 a \wedge \star a,
\end{equation}
where we introduced the matrix $\Xi^0$, whose component is given by $\Xi^0_{ab} \equiv \delta_a^1 \delta_b^1$ in our gauge.
Here we neglected the constant term $V(f_\pi^2)$ resulting from the potential.
The condensation of $z$ giving a mass to the dynamical $\U(1)$ gauge field $a$
via the Higgs mechanism.
Then, we can safely perform the integration over $a$ by using the equation of motion:
\begin{equation}
 a = - \rmi \tr (\Xi^0 U^\dag \diff U) + O (\diff^3). 
  \label{eq: eom}
\end{equation}
As a consequence, we derive the effective Lagrangian for the N\'eel phase as
\begin{equation}
 \begin{split}
  \mL_{\mathrm{N\acute{e}el}}
  &= f_\pi^2 
  \left[\tr 
  ( \Xi^0 \diff U^\dag \wedge \star \diff U )
  +  \tr (\Xi^0 U^\dag \diff U) \wedge \star \tr (\Xi^0 U^\dag \diff U)
  \right]
  \\
  &= \frac{f_\pi^2}{2} \tr 
  (\diff \Xi \wedge \star \diff \Xi ),
 \end{split}
   \label{eq:Eff-Lag-Neel}
\end{equation}
where, in the second line, we introduced the matrix field $\Xi(x) = U (x) \Xi^0 U^\+(x)$.
% whose component is given by
%\begin{equation}
% \Xi_{ab} (x) \equiv U (x) \Xi^0_{ab} U(x).
%\end{equation}
Eq.~\eqref{eq:Eff-Lag-Neel} gives an effective Lagrangian for the N\'eel phase 
describing the dynamics of the Nambu-Goldstone mode~(see e.g., Ref.~\cite{Bar:2003ip}).
Since the flavor symmetry $g \in \PSU(N)_F$ acts on $\Xi $ as $\Xi \to g \Xi g^{-1}$, 
the effective Lagrangian \eqref{eq:Eff-Lag-Neel} respects the $\PSU(N)_F$ symmetry manifestly.

Based on the constructed effective theory, we then confirm the 't Hooft anomaly matching in the N\'eel phase. 
A key question arising from the bottom-up view of the low-energy effective theory \eqref{eq:Eff-Lag-Neel} is whether we have a symmetry corresponding to $\U(1)_M^{D-3}$ within its framework.
At first glance, the answer seems negative since there is no corresponding symmetry
in the effective Lagrangian \eqref{eq:Eff-Lag-Neel}.
However, the crucial observation is 
the second homotopy group of the coset is nontrivial:
%that symmetry breaking pattern results in a nontrivial second homotopy group
\begin{align}
 \pi_2 \left( \frac{\U(N)}{\U(1) \times \U(N-1)} \right)
 = \pi_2 (\CP^{N-1}) = \Z.
\end{align}
This topological number allows the N\'eel phase to possess topologically stable configurations.
As a result, we have the topologically conserved $(D-2)$-form current:
\begin{align}
\star J_\mathrm{S} 
 = - \frac{\rmi}{2 \pi} \tr (\Xi_0 \diff U^{-1}\wedge \diff U)
 = - \frac{\rmi}{2 \pi} \tr (\Xi \diff \Xi \wedge \diff \Xi),
\end{align}
whose coefficient is chosen so that $\int_{S^2} J_\mathrm{S} \in \Z$ with a spatial $2$-sphere $S^2$.
This $(D-2)$-form current is referred to as the Skyrmion current since the associated charge in the $(2+1)$d case indeed counts a number of the magnetic (or baby) Skyrmions. 
We then identify the Skyrmion current as the $(D-2)$-form magnetic current in the original $\CP^{N-1}$ model.
On the other hand,
from the top-down viewpoint, this identification is naturally justified because we can rewrite the magnetic current using Eq.~\eqref{eq: eom} as
\begin{align}
 \star J_{\mathrm{M}} =
 \frac{1}{2 \pi} \diff a 
 = - \frac{\rmi}{2 \pi} \tr ( \Xi_0 \diff U^{-1} \wedge \diff U)
 = \star J_\mathrm{S} .
\end{align}

We thus find that the low-energy effective theory of the N\'eel phase realizes the $\U(1)_M^{[D-3]}$ symmetry as the topological symmetry. 
By inserting the background $(D-2)$-form gauge field $K$, the effective action acquires the additional coupling as
\begin{align}
 S_{\mathrm{N\acute{e}el}}  [\Xi;K]
 = \frac{f_\pi^2}{2} \int \tr (\diff \Xi \wedge \star \diff \Xi) 
 + \rmi \int K \wedge \star J_\mathrm{S}.
\end{align}
In the following, we will show that the additional term actually saturates both the $\PSU(N)_F \times \U(1)^{[D-3]}_M$ and $\mR \times \U(1)^{[D-3]}_M$ anomalies.

\paragraph{$\PSU(N)_F \times \U(1)^{[D-3]}_M$ anomaly matching.}
Let us first turn on the $\PSU(N)_F$ background gauge field.
This just replaces the partial derivatives in the effective Lagrangian with the covariant ones.
However, this na\"ive replacement violates the conservation of the Skyrmion current.
The violation is canceled if one modifies the Skyrmion current by adding the gauge-invariant term:
\begin{align}
 \star J_\mathrm{S} 
 = - \frac{\rmi}{2 \pi} \tr (\Xi_0 D_A U^{-1}\wedge  D_A U )
 - \frac{1}{2 \pi} \tr (\Xi_0 U^{-1} F[A] U)
 = - \frac{\rmi}{2 \pi} \tr (\Xi_0 \diff [ U^{-1} D_A U ] ),
 \label{eq:GW-current}
\end{align}
where we defined the covariant derivative of the coset
$D_A U \equiv (\diff - \rmi A) U $.
The modified current \eqref{eq:GW-current} is the so-called Goldstone-Wilczek current~\cite{Goldstone:1981kk,Bar:2003ip,Wiese:2005tg} (see also Ref.~\cite{Jackiw:1975fn} for a related early work).
Although the second term fixes the conservation law or the $\U(1)_M^{[D-3]}$ symmetry, 
it, instead, spoils the $\Z_N$ $1$-form gauge invariance. 

On the other hand, we can recover the $\Z_N$ $1$-form gauge invariance by adding $B/2\pi$ with the $2$-form $\Z_N$ gauge field $B$, which yields another definition of the Skyrmion current:
\begin{equation}
 \begin{split}
  \star J_\mathrm{S} 
  &= -\frac{\rmi}{2 \pi} \tr (\Xi_0 D_A U^{-1}\wedge  D_A U )
  - \frac{1}{2 \pi} \tr (\Xi_0 U^{-1} F[A] U) +\frac{B}{2 \pi}
  \\
  &= -\frac{\rmi}{2 \pi} \tr (\Xi D_A \Xi \wedge  D_A \Xi )
  - \frac{1}{2 \pi} \tr (\Xi F[A] ) +\frac{B}{2 \pi}.
 \end{split}
\label{eq: inv skyr curr}
\end{equation}
Nevertheless, this modified current now spoils the $\U(1)_M^{[D-3]}$ large gauge invariance due to the fractional quantization coming from the last term.
This competing behavior precisely reflects the mixed $\PSU(N)_F \times \U(1)^{[D-3]}_M$ anomaly.
Therefore, we have shown the $\PSU(N)_F \times \U(1)^{[D-3]}_M$ anomaly in the N\'eel phase.

\paragraph{$\mR \times \U(1)^{[D-3]}_M$ anomaly matching.}
We next investigate the $\mR \times \U(1)^{[D-3]}_M$ anomaly in the N\'eel phase
by putting the model on a nonorientable manifold, say $\mM_D =\mM_{D-2} \times \RP^2$.
Recall Eq.~\eqref{eq: cp anom Ddim}, where the magnetic current takes the half-integer value on the manifold.
The equation of motion~\eqref{eq: eom} tells us
\begin{align}
 \int w_2 
 = \frac{1}{2 \pi} \int \diff a 
 = - \frac{\rmi}{2 \pi} \int \tr ( \Xi \diff \Xi \wedge \diff \Xi).
\end{align}
In other words, a half Skyrmion shows up on the nonorientable manifold
and again breaks the $\U(1)^{[D-3]}_M$ large gauge invariance.
Therefore, thanks to the topological current, the N\'eel phase is shown
to be consistent with the $\mR \times \U(1)^{[D-3]}_M$ anomaly matching.

\subsection{$\U(1)$ spin liquid phase}~\label{sec: u1 spin}
In this subsection, we examine the anomaly matching in the $\U(1)$ spin liquid phase, where the $\U(1)_M^{[D-3]}$ magnetic symmetry is spontaneously broken.
The discussion in this subsection also serves as a basis to discuss the anomaly matching in the VBS phase.

%As discussed in the beginning of this section, the magnetic symmetry $\U(1)^{[D-3]}_M$ is spontaneously broken in the $\U(1)$ spin liquid phase. Instead, 
In the $\U(1)$ spin liquid phase, the $N$-component complex scalar field is massive, and we can simply integrate out it in the original $\CP^{N-1}$ Lagrangian \eqref{eq: cpn model}. 
As a result, the system only contains the dynamical $\U(1)$ gauge field $a$ in the low-energy limit, which results in the simple effective Lagrangian
\begin{equation}
 \mL_{\mathrm{\U(1)SL}} = \frac{1}{2g^2} \diff a \wedge \star \diff a.
  \label{eq:Eff-Lag-SL1}
\end{equation}
This rather trivial effective Lagrangian is, of course, consistent with a general symmetry-based argument.
Besides, the S-duality maps the Maxwell theory onto
\begin{equation}
 \widetilde{\mL}_{\mathrm{\U(1)SL}} 
  = \frac{1}{2 \tilde{g}^2} \diff \tilde{a} \wedge \star \diff \tilde{a},
  \label{eq:Eff-Lag-SL2}
\end{equation}
where $g \tilde{g} = 2\pi$ and $\tilde{a}$ is a $(D-3)$-form $\U(1)$ gauge field which is dual to $a$.

In general, when a continuous $p$-form symmetry is spontaneously broken,
an associated Nambu-Goldstone boson is described by a $p$-form gauge field~\cite{Gaiotto:2014kfa, Lake:2018dqm}.
Since $\U(1)^{[D-3]}_M$ is broken in the $\U(1)$ spin liquid phase, 
we can identify $\tilde{a}$ as the Nambu-Goldstone boson associated with the $\U(1)^{[D-3]}_M$ symmetry breaking.
In the following, we will mainly use the dual gauge field to examine its symmetries and anomaly matching.

% In general, when a continuous $p$-form symmetry is spontaneously broken,
% an associated Nambu-Goldstone boson is described by a $p$-form gauge field~\cite{Gaiotto:2014kfa, Lake:2018dqm}.
% In our case, the $\U(1)$ spin liquid phase inevitably supports the $(D-3)$-form gauge field $\tilde{a}$, whose dynamics is captured by the following effective Lagrangian
% \begin{equation}
%  \widetilde{\mL}_{\mathrm{\U(1)SL}} 
%   = \frac{1}{2 \tilde{g}^2} \diff \tilde{a} \wedge \star \diff \tilde{a}. 
%   \label{eq:Eff-Lag-SL2}
% \end{equation}
% Thanks to the abelian duality, or the S-duality, we know that these two theories defined in Eqs.~\eqref{eq:Eff-Lag-SL1} and \eqref{eq:Eff-Lag-SL2} are indeed equivalent with $g \tilde{g} = 2\pi$; that is, the Nambu-Goldstone mode $\tilde{a}$ represents the dual gauge field of the original gauge field $a$, and vice versa.
% In the following, we will mainly use the dual gauge field to examine symmetry and anomaly matching.

Let us take a closer look at the symmetry structure of the above low-energy effective theory.
In addition to the $\U(1)_M^{[D-3]}$ symmetry generating the shift of the Nambu-Goldstone boson $\tilde{a}$ as $\tilde{a} \to \tilde{a} + \theta$, 
one easily finds the low-energy effective Maxwell theory enjoys the emergent $1$-form shift symmetry for the original gauge field $a$, which we call the $\U(1)_E^{[1]}$ symmetry. 
The Noether current for the $\U(1)_E^{[1]}$ symmetry is simply given by 
\begin{equation}
 \star J_{\mathrm{E}} = \frac{1}{2\pi} \diff \tilde{a},
  \label{eq: VBS vortex}
\end{equation}
which is conserved due to the absence of electrically charged objects in the spin liquid phase.
The current $\star J_E$ characterizes line-like defects in this phase (e.g., a vortex in $D=3$).

The vital point for the subsequent discussion is that the Maxwell theory suffers from the mixed anomaly between the $\U(1)_M^{[D-3]}$ and $\U(1)_E^{[1]}$ symmetries.
This observation motivates us to identify a discrete counterpart of the $\U(1)_E^{[1]}$ symmetry, or the $(\Z_N^{[1]})_E$ symmetry, as a remnant of the $\PSU(N)_F$ symmetry (recall that the original $1$-form gauge transformation in Eq.~\eqref{eq: u1 1form} is consistent with the above identification).
We first note that gauging the $\U(1)_M^{[D-3]}$ symmetry 
is accomplished by replacing $\diff \tilde{a}$ with $\diff \tilde a - K$, an invariant combination under the local $\U(1)_M^{[D-3]}$ transformation:
\begin{align}
 \tilde{a}(x) \ra \tilde{a}(x) + \theta(x),
 \quad 
 K(x) \ra K(x) + \diff \theta(x).
 \label{eq:Mag-gauge-tr}
\end{align}
Furthermore, one can introduce the $2$-form $\Z_N$ gauge field $B$ through the minimal coupling to the current in Eq.~\eqref{eq: VBS vortex}.
We eventually find gauging the $\U(1)_M^{[D-3]}$ and $(\Z_N^{[1]})_E$ symmetries leads to the following effective action:
\begin{equation} 
 \begin{split}
  &\widetilde{S}_{\mathrm{\U(1)SL}} [\tilde{a};K,B]
  = \int \frac{1}{2 \tilde{g}^2} (\diff \tilde{a} - K) 
  \wedge \star (\diff \tilde{a} - K)
  + \frac{\rmi}{2\pi} \int B \wedge \diff \tilde{a},
  % \\
  % \leftrightarrow
  % &~S_{\mathrm{\U(1)SL}} [a;K,B]
  % = \int \frac{\tilde{g}^2}{8 \pi^2} \diff a \wedge \star \diff a
  % + \int \frac{\rmi}{2 \pi} K \wedge \diff a ,
 \end{split}
 \label{eq: u1 eft1}
\end{equation}
where $B$ obeys the fractional quantization \eqref{eq: b quant}.

The second term in Eq.~\eqref{eq: u1 eft1} enables us to confirm that the 't Hooft anomaly is consistently matched in the $\U(1)$ spin liquid phase.
In fact, this term respects the $\Z_N$ $1$-form gauge invariance, but spoils the $\U(1)^{[D-3]}_M$ gauge invariance under Eq.~\eqref{eq:Mag-gauge-tr} as
\begin{align}
 \delta_\theta \widetilde{S}_{\mathrm{\U(1)SL}} [\tilde{a};K,B]
 = \frac{\rmi}{2 \pi} \int B \wedge \diff \theta.
\end{align}
This violation is exactly the same as Eq.~\eqref{eq: psu anom Ddim}
and properly reproduces the phase shift attached to the $\PSU(N)_F \times \U(1)^{[D-3]}_M$ anomaly.
Moreover, the same discussion works for the $\mR \times \U(1)^{[D-3]}_M$ anomaly.
On the nonorientable manifold, the effective action is shown to take the following form:
\begin{align}
  \widetilde{S}_{\mathrm{\U(1)SL}} [\tilde{a};K,w_2]
 = \int \frac{1}{2 \tilde{g}^2} (\diff \tilde{a} - K) 
 \wedge \star (\diff \tilde{a} - K)
 + \frac{\rmi}{2 \pi} \int \pi w_2 \wedge \diff \tilde{a},
\end{align}
which consistently matches the $\mR \times \U(1)^{[D-3]}_M$ anomaly.
In both cases, the 't Hooft anomalies are saturated by the discrete counterparts of the $\U(1)_E^{[1]}$ symmetry and the current~\eqref{eq: VBS vortex} carrying nontrivial charge under the $\PSU(N)_F$ and $\mR$ symmetry.

\subsection{Valence bond solid (VBS) phase}~\label{sec: vbs}

Based on the discussion in the previous subsection, 
we consider the effect of the charge-$n$ monopole deformation of the theory and discuss the anomaly matching in the VBS phase.

The valence bond solid phase appears when the charge-$n$ monopole condenses so that the $(\Z_n^{[D-3]})_M$ symmetry is spontaneously broken.
As discussed above, identifying the discrete subgroup of the electric $1$-form symmetry, or the $(\Z_N^{[1]})_E$ symmetry, as a remnant of the $\PSU(N)_F$ symmetry, we can apply the same discussion as the $\U(1)$ spin liquid phase;
the effective Lagrangian in the VBS phase also contains the followings:
\begin{align}~\label{eq: vbs anom}
 S_{\mathrm{anom}}^{\mathrm{VBS}} [\tilde{a};B,w_2]
 = \frac{\rmi}{2 \pi} \int B \wedge \diff \tilde{a} 
 + \frac{\rmi}{2 \pi} \int \pi w_2 \wedge \diff \tilde{a}
\end{align}
in the presence of the background gauge fields for the $\PSU(N)_F$ and $\mR$ symmetries.
We thus again see these terms possibly spoil the $(\Z_n^{[D-3]})_M$ gauge invariance depending on the value of the monopole charge (see Table.~\ref{tbl: anom w monopole}), saturating our 't Hooft anomalies.
In the remaining of this subsection, we will show the consequence of the anomaly matching in the VBS phase: deconfined excitations localized on a domain wall in $D=3$ and topological order in $D=4$.

\paragraph{Anomaly inflow for $3$d VBS domain wall.}

Let us first consider the $3$d VBS phase, where the dual gauge field $\tilde{a}$ denotes a scalar field~\cite{Komargodski:2017smk}.
We find that the effective theory equipped with the $(\Z_n)_M$ shift symmetry of the dual scalar $\tilde{a}$ is given by the sine-Gordon model with a potential $V (\tilde{a}) = \rho_M^2 (1 - \cos n \tilde{a})$. 
The cosine potential typically drives the system to break the $(\Z_n)_M$ symmetry spontaneously, resulting in the VBS phase.
Turning on the $(\Z_n)_M$ gauge fields $K_n$ and $H$, we obtain the gauged effective action of the VBS phase as
\begin{align}
 S_{\mathrm{VBS}}^{3\mathrm{D}} 
 [\tilde{a};K_n,H]
 = \frac{1}{2 \tilde{g}^2} (\diff \tilde{a} - K_n) 
 \wedge \star (\diff \tilde{a} - K_n)
 + \rho^2_M 
 \big[ 1 - \cos (n \tilde{a} - H) \big].
\end{align}
Compared with the $\U(1)$ spin liquid phase, the $3$d VBS phase can have the domain-wall solution due to a $n$-fold degeneracy of the classical vacua at  
$\tilde{a} \equiv 2 \pi k/n\, (k=0,1,\cdots, n-1)$.
As discussed in Ref.~\cite{Komargodski:2017smk},
a nontrivial excitation is localized on a domain wall in the VBS phase with the $\PSU(N) \times (\Z_n)_M$ anomaly due to the anomaly inflow. 
%when the VBS phase possessing the 't Hooft anomaly realizes the domain-wall solution, there is a nontrivial excitation localized on the domain wall due to the anomaly inflow attached to the $\PSU(N) \times (\Z_n)_M$ anomaly.
Let us confirm their discussion also respects the $\mR \times (\Z_n)_M$ anomaly.

For simplicity, let us consider the $n = 2$ case, but generalization with arbitrary $n$ is straightforward.
The cosine potential pins the scalar field $\tilde{a}$ to $\tilde{a}= 0,~\pi$ leading to $2$-fold degeneracy of the classical vacua.
Let us then consider a domain-wall solution interpolating two vacua along e.g., the $x^3$-direction by imposing the boundary condition%
\footnote{
Employing this boundary condition could be also regarded as an insertion of a nontrivial flux $H$ with $\int_{x^3} \diff H \in 2 \pi \Z$
%$H = 2 \pi \Theta(x^3)$ 
for the background $\Z_2$ gauge field on a three-dimensional torus.
}: $\tilde{a} (x) = 0$ at $ x^3 = - \infty$ and $\tilde{a}(x) = \pi$ at $ x^3 = \infty$.
Neglecting a thickness of the realized domain wall, we simply put such a solution as $\tilde{a} (x) = \pi \Theta(x^3)$ with a Heaviside step function $\Theta (x)$.
Substituting this solution into the anomalous part of the effective action~\eqref{eq: vbs anom}, we obtain the following result nonvanishing on the wall:
\begin{align} \label{eq: wall anom}
 S^{\mathrm{ano}}_{\mathrm{VBS}}
 [\tilde{a};B,w_2] \big|_{\tilde{a} = \pi \Theta (x^3)}
 = \frac{\rmi}{2} \int_{x^1,x^2} B(x^1,x^2,0)
 + \frac{\rmi}{2} \int_{x^1,x^2} \pi w_2(x^1,x^2,0).
\end{align}
Because of their coefficients, they don't satisfy the $\Z_2$ $1$-form gauge invariance for $B$ and the $2 \pi$ periodicity for $\pi w_2$ on the wall and seem ill-defined itself.

%This apparent contradiction is resolved by letting the theory on the wall be an anomalous two-dimensional one so that the total system is consistent.
This apparent contradiction is resolved by letting the theory on the wall have two-dimensional anomalies so that the total system is consistent.
Such an anomalous system which cancels both the anomalous terms in Eq.~\eqref{eq: wall anom} is the two-dimensional $\CP^{N-1}$ model with a $\theta$ term at $\theta = \pi$.
At this $\theta$ angle, this $2$d model has the charge-conjugation symmetry $\mC$
and also possesses the $\PSU(N)_F \times \mC$~\cite{Komargodski:2017smk, Komargodski:2017dmc}
and $\mR \times \mC$ anomalies~\cite{Sulejmanpasic:2018upi}.
In Ref.~\cite{Komargodski:2017smk}, it is discussed that $\PSU(N)_F \times \mC$ anomaly saturates the first anomalous term on the wall in Eq.~\eqref{eq: wall anom}.
It is also pointed out the existence of an excitation carrying a $\PSU(N)_F$ quantum number, which is deconfined on the wall but confined in the bulk.
Here, we find that the $\mR \times \mC$ anomaly in the $2$d $\CP^{N-1}$ model also saturates the second anomalous term in Eq.~\eqref{eq: wall anom}.
Therefore, in the presence of the $\mR \times (\Z_n)_M$ anomaly,
the deconfined excitation on the wall must transform as in Eq.~\eqref{eq: reflection} under the reflection symmetry.

\paragraph{Topological order in $4$d (or higher-dimensional) VBS phase.}

In $D \ge 4$, a charge-$n$ Higgs field plays a role analogous to the cosine potential in $D=3$, which breaks the $\U(1)^{[D-3]}_M$ symmetry down to $(\Z_n^{[D-3]})_M$ (recall the discussion on the monopole deformation of the theory given in Sec.~\ref{sec: global}). 
The gauged effective action is thus identified as 
\begin{align}
 S_{\mathrm{VBS}}^{D\geq 4} [\tilde{a}, \eta ;K_n, H]
 = \int \frac{1}{ 2 \tilde{g}^2} (\diff \tilde{a} - K_n) \wedge \star (\diff \tilde{a} - K_n)
+\frac{\rho^2_M}{2} (\diff \eta - n \tilde{a} + H) \wedge \star (\diff \eta - n \tilde{a} + H),
 \label{eq:VBS-Higgs-gauged}
\end{align}
where $\eta$ is a $(D-4)$-form $\U(1)$ gauge field.
In the spacetime dimension higher than $3$ (or $D \ge 4$), the $(\Z_n^{[D-3]})_M$ symmetry is spontaneously broken when $\rho^2_M$ is relevant.
In that case, we can show that the low-energy dynamics is governed by a topological quantum field theory, known as the BF theory~\cite{Bergeron:1994ym, Hansson:2004wca, Cho:2010rk, Chen:2015gma, Putrov:2016qdo, Hidaka:2019jtv}.
To show this, we
first rewrite the effective action \eqref{eq:VBS-Higgs-gauged} with the help of an auxiliary $3$-form field $h$ as 
\begin{equation}
 S_{\mathrm{VBS}}^{D\geq 4} [\tilde{a}, \eta ;K_n, H]
  = \int
  \frac{1}{ 2 \tilde{g}^2} (\diff \tilde{a} - K_n) 
  \wedge \star ( \diff \tilde{a} - K_n )
  + \frac{1}{8\pi^2 \rho_M^2} h \wedge \star h
  - \frac{\rmi}{2\pi} h \wedge ( \diff \eta - n \tilde{a} + H ).
\end{equation}
Instead of integrating out $h$, which reproduces the original effective Lagrangian \eqref{eq:VBS-Higgs-gauged}, we now integrate out $\eta$ by using its equation of motion $\diff h = 0$, which can be solved as $h = \diff b$ with a $2$-form field $b$.
As a consequence, we obtain the low-energy effective action for the VBS phase in $D\geq 4$  given by  
\begin{align}
 S_{\mathrm{top}} [\tilde{a},b;K_n,H]
 =  - \frac{\rmi n}{2 \pi} \int b \wedge (\diff \tilde{a} - K_n ),
 \label{eq:Top-order}
\end{align}
where we neglected the higher derivative terms.
The BF theory defined by the effective action \eqref{eq:Top-order} describes a $\Z_n^{[D-3]}$ topological order such as a BCS superconductor~\cite{Hansson:2004wca}.
In addition to the original $\Z_n^{[D-3]}$ magnetic symmetry, the BF theory possesses an emergent $\Z_n^{[2]}$ symmetry acting on $b$ as 
\begin{equation}
  b  \to b + \frac{1}{n} \diff \lambda \quad \mathrm{with} \quad 
   \int_{\Sigma_2} \diff \lambda \in 2\pi \Z.
\end{equation}
Nevertheless, one finds that the introduction of the $(\Z_n^{[D-3]})_M$ background gauge field spoils this emergent symmetry as
\begin{align}
 \delta_\lambda S_{\mathrm{top}} 
 =  - \rmi \int \frac{\diff \lambda}{2\pi} \wedge \diff \tilde{a}
 + \rmi \int \frac{\diff \lambda}{2\pi} \wedge K_n 
 \in \frac{2\pi \rmi \Z }{n},
 \label{eq:Top-order2}
\end{align}
where we used a constraint $n K_n = \diff H$ associated with the $(\Z_n^{[D-3]})_M$ gauge field.
This shows the presence of the mixed anomaly between the $(\Z_n^{[D-3]})_M$ and $\Z_n^{[2]}$ symmetries, and implies spontaneous $(\Z_n^{[D-3]})_M$ symmetry breaking in the VBS phase.
One can directly show the ground state degeneracy of the BF theory \eqref{eq:Top-order} depends on the topology of the spatial manifold (see e.g., Refs.~\cite{Bergeron:1994ym, Hansson:2004wca, Cho:2010rk, Chen:2015gma, Putrov:2016qdo, Hidaka:2019jtv} for a detailed discussion).

\section{Application: $4$d anomaly at finite temperature}~\label{sec: 4d anom}
As an application of the anomaly matching, we shall discuss the fate of the $\PSU(N)_F \times \U(1)^{[1]}_M$ and $\mR \times \U(1)^{[1]}_M$ anomalies of the four-dimensional $\CP^{N-1}$ model at finite temperature and apply it to constrain its possible phase diagram.

\subsubsection{Persistence of the anomaly at finite temperature}
A usual symmetry, or a $0$-form symmetry, acts on the point-like local operator, and the associated 't Hooft anomaly usually disappears when we consider a finite-temperature situation (see, however, Ref.~\cite{Tanizaki:2017qhf} for a way to make a $0$-form anomaly to survive under circle compactification).
Nevertheless, since the anomalies for the $4$d $\CP^{N-1}$ model involve the $1$-form magnetic symmetry, which acts on the line operator, they survive and forbid the model from being a trivial phase even at finite temperature.
In this subsection, after demonstrating the persistence of the $\PSU(N)_F \times \U(1)^{[1]}_M$ and $\mR \times \U(1)^{[1]}_M$ anomalies at finite temperature, we will show the consequence of anomaly matching. 
We also explicitly show the large-$N$ phase diagram 
of the $\CP^{N-1}$ \textit{nonlinear} sigma model is consistent with the anomalies.

First of all, we explain that the anomalies in the $4$d $\CP^{N-1}$ model indeed survive at finite temperature.
Suppose that our system lives on $\mM_4 = S^1 \times \mM_3$ and regard the circumstance in the $x^1$ direction as the inverse temperature $1/T$.
Then, the magnetic symmetry in four dimensions splits into $1$-form and $0$-form symmetries, which act on the spacial magnetic loop and the point-like object, respectively.
In the four-dimensional language,
this can be seen as the following decomposition of the $\U(1)^{[1]}_M$ gauge field $K$:  
\begin{align}
 K = K^{(2)} + T \diff x^1 \wedge K^{(1)},
\end{align}
where $K^{(2)}$ and $K^{(1)}$ are $2$-form and $1$-form gauge fields on the 
three-dimensional spacial manifold $\mM_3$.
Note that we can always introduce this specific configuration at any temperature.
%Corresponding to this, we also need to decompose the $2$-form $\Z_N$ gauge field $B$ for the $\PSU(N)_F$ symmetry as
%\begin{align}
%B = B^{(2)} + T \diff x^1 \wedge B^{(1)}.
%\end{align}
%Likewise, $B^{(2)}$ and $B^{(1)}$ are $2$-form and $1$-form $\Z_N$ gauge fields on the spacial manifold $\mM_3$.
In this finite-temperature setup, the magnetic $\U(1)_M^{[1]}$ gauge transformation given by $ K \to K + \diff \theta$ can be decomposed into
\begin{align}
 \diff \theta = \diff \theta^{(2)} + T \diff x^1 \wedge \diff \theta^{(1)},
 \label{eq:decomp-dtheta}
\end{align}
where $\theta^{(2)}$ and $\theta^{(1)}$ are the gauge parameters for the $1$-form and $0$-form magnetic symmetries, respectively.
Note that this decomposition is equivalent to the separate gauge transformations
$ K^{(2)} \ra K^{(2)} + \diff \theta^{(2)},~
 K^{(1)} \ra K^{(1)} + \diff \theta^{(1)}$.
On the other hand, the $2$-form gauge field $B$ just reduces $B^{(2)}$, a $2$-form gauge field on $\mM_3$%
\footnote{This is because this gauge field is associated with the quotient part of the $0$-form symmetry $\PSU(N)_F$.
However, if we take a twisted boundary condition by $\PSU(N)_F$ in the $x^1$ direction, this yields a new $\Z_N$ symmetry and allows us to introduce a $\Z_N$ $1$-form gauge field $B^{(1)} $analogous to $K^{(1)}$~\cite{Tanizaki:2017qhf}.
However, we don't consider the special boundary condition here and focus on the compactification with the periodic one.}.
Substituting the above decomposition of $K$ and the reduced expression of $B$ into Eq.~\eqref{eq: psu anom Ddim} and 
performing the imaginary-time integral, we find the partition function for the $4$d $\CP^{N-1}$ model acquires the anomalous phase as 
\begin{align}
 Z[K + \diff \theta ,A,B] = Z[K,A,B] 
 \exp \left( - \frac{ \rmi }{2 \pi} \int_{\mM_3}
 %\left[ 
 \diff \theta^{(1)} \wedge B^{(2)}
% + \diff \theta^{(2)} \wedge B^{(1)} \right]
  \right).
\end{align}
The first takes exactly the same form as the $\PSU(N)_F \times \U(1)_M$ anomaly in the $3$d $\CP^{N-1}$ model.%,
%while the last term represents the $3$d mixed anomaly between the $\Z_N$ and $\U(1)^{[1]}_M$ symmetries
%\footnote{This anomaly is saturated by a theory with a real scalar field $\varphi$ with a $\Z_N$ shift symmetry described by e.g.
%\begin{align}
% \frac{\rho^2}{2}(\diff \varphi - B^{(1)})\wedge \star (\diff \varphi - B^{(1)})
% + \frac{\rmi}{2 \pi} K^{(2)} \wedge (\diff \varphi - B^{(1)})
% +\lambda [1- \cos (N \varphi - C^{(0)}) ],
%\end{align}
%where $C^{(0)}$ is given by the trace of the time-component of the $U(N)$ gauge field $A$.
%The scalar field $\varphi$ may be thought of as a Wilson loop wrapping the time direction after the reduction (i.e., $\varphi = \int_{S^1}a$).
%}.
In addition, when the spatial manifold is given by 
$\mM_3 = S^1 \times \RP^2$,
Eq.~\eqref{eq: cp anom Ddim} with the 
decomposition \eqref{eq:decomp-dtheta} results in the anomalous phase factor of the partition function:
\begin{equation}
 Z[K + \diff \theta, w_2] = Z[K, w_2] 
  \exp 
  \left( - \frac{\rmi}{2\pi} \int_{\mM_3} \diff \theta^{(1)} \wedge \pi w_2 \right).
\end{equation}
This matches the $\mR \times \U(1)_M$ anomaly in the $3$d $\CP^{N-1}$ model.
The same discussion also works in the presence of the monopole deformation, and 
the finite-temperature $\CP^{N-1}$ model possesses the $\PSU(N)_F \times (\Z_n)_M$ and $\mR \times (\Z_n)_M$ anomalies as the zero-temperature one.

It is worthwhile emphasizing that both the $\PSU(N)_F \times \U(1)^{[1]}_M$ and $\mR \times \U(1)^{[1]}_M$ mixed anomalies survive at any temperature.
This indicates that the system still stays in the nontrivial phase even in the high-temperature limit, where the imaginary-time direction is not visible, and the spacetime manifold effectively reduces to the $3$d one.
However, it is also important to note that the $\CP^{N-1}$ model is a cutoff theory with an ultraviolet cutoff
---in particular, one can manifestly see the nonlinear one as a cutoff theory.
As a result, at higher temperature than the ultraviolet cutoff, 
the field theoretical description will break down, 
and we do not know the fate of the 't Hooft anomalies.
This property directly affects our subsequent discussion on the finite-temperature phase diagram of the $\CP^{N-1}$ model.

\subsubsection{Possible phase structure}
Based on the persistence of the 't Hooft anomalies in the finite-temperature $\CP^{N-1}$ model, we discuss
its consequence to the phase diagram and thermal phase transition.
Here we present possible simple phase structures consistent with the anomalies 
by restricting ourselves to the symmetry breaking scenarios discussed in Sec.~\ref{sec: matching}.

While the system cannot reside in the trivial phase in the regime where the $\CP^{N-1}$ model description is applicable, 
we need to take care of the fact the $4$d $\CP^{N-1}$ model is a field theory formulated with a UV cutoff denoted by $\Lambda_{\mathrm{cutoff}}$ (e.g., the inverse lattice constant of spin systems).
Above the cutoff, the $\CP^{N-1}$ model fails to describe underlying quantum many-body systems, and our anomalies and anomaly matching argument could break down (see also the discussion given in Sec.~\ref{sec: summary}).

\begin{figure}[t  ]%
 \centering
 \subfloat[]{\includegraphics[scale=.27]{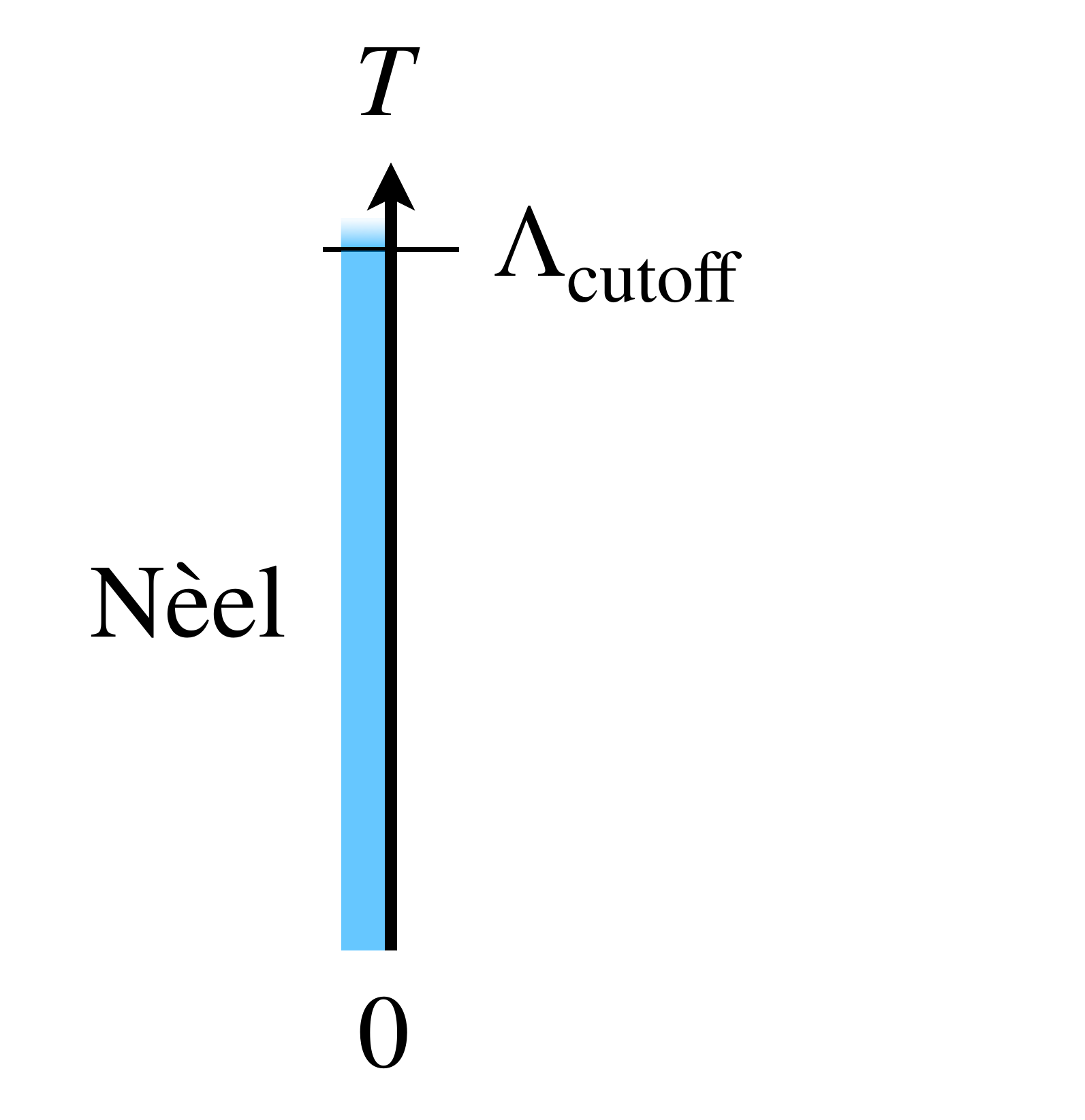}\label{fig: neel} }
 \subfloat[]{\includegraphics[scale=.27]{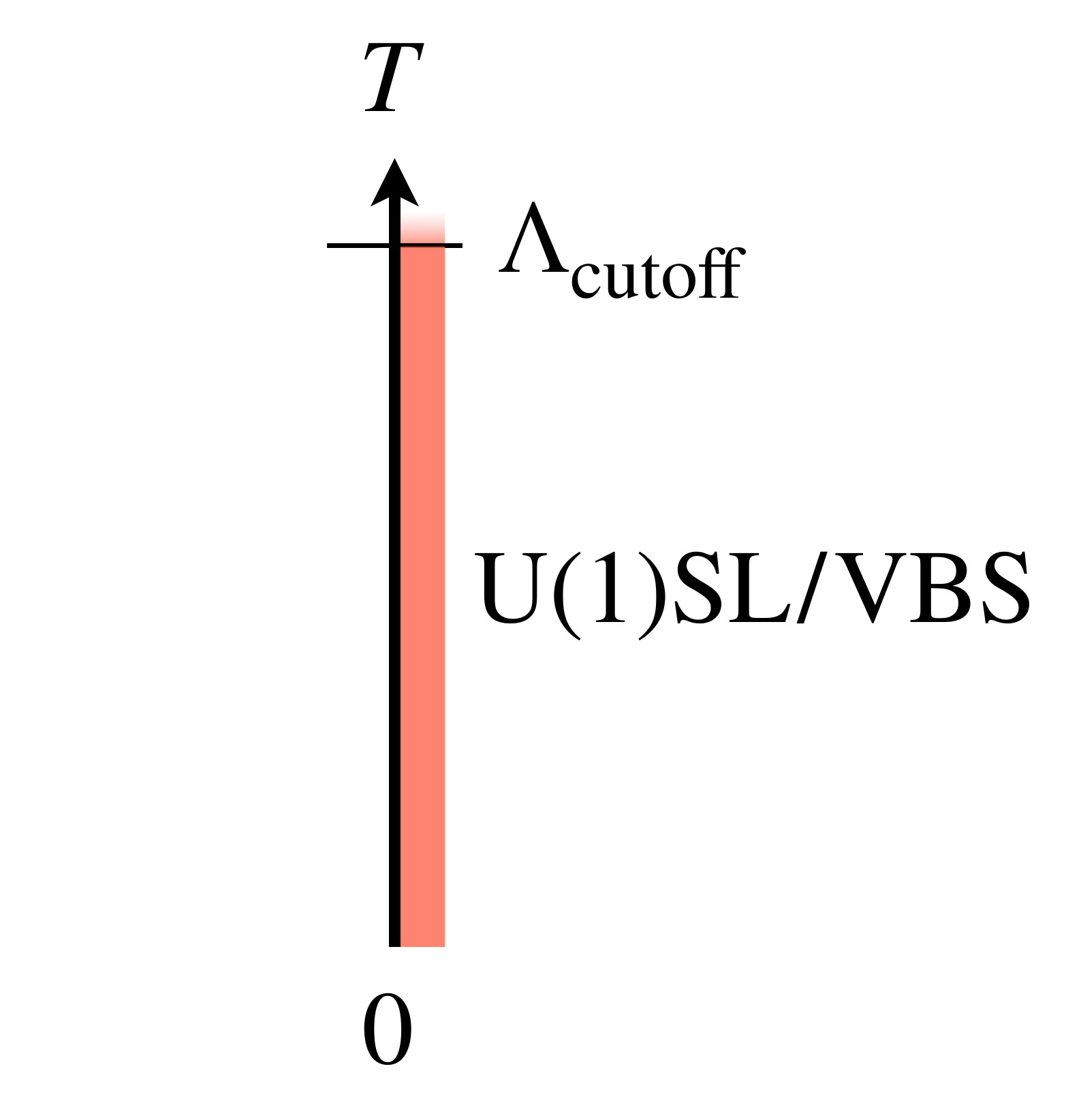}\label{fig: u1sl} }
 \subfloat[]{\includegraphics[scale=.27]{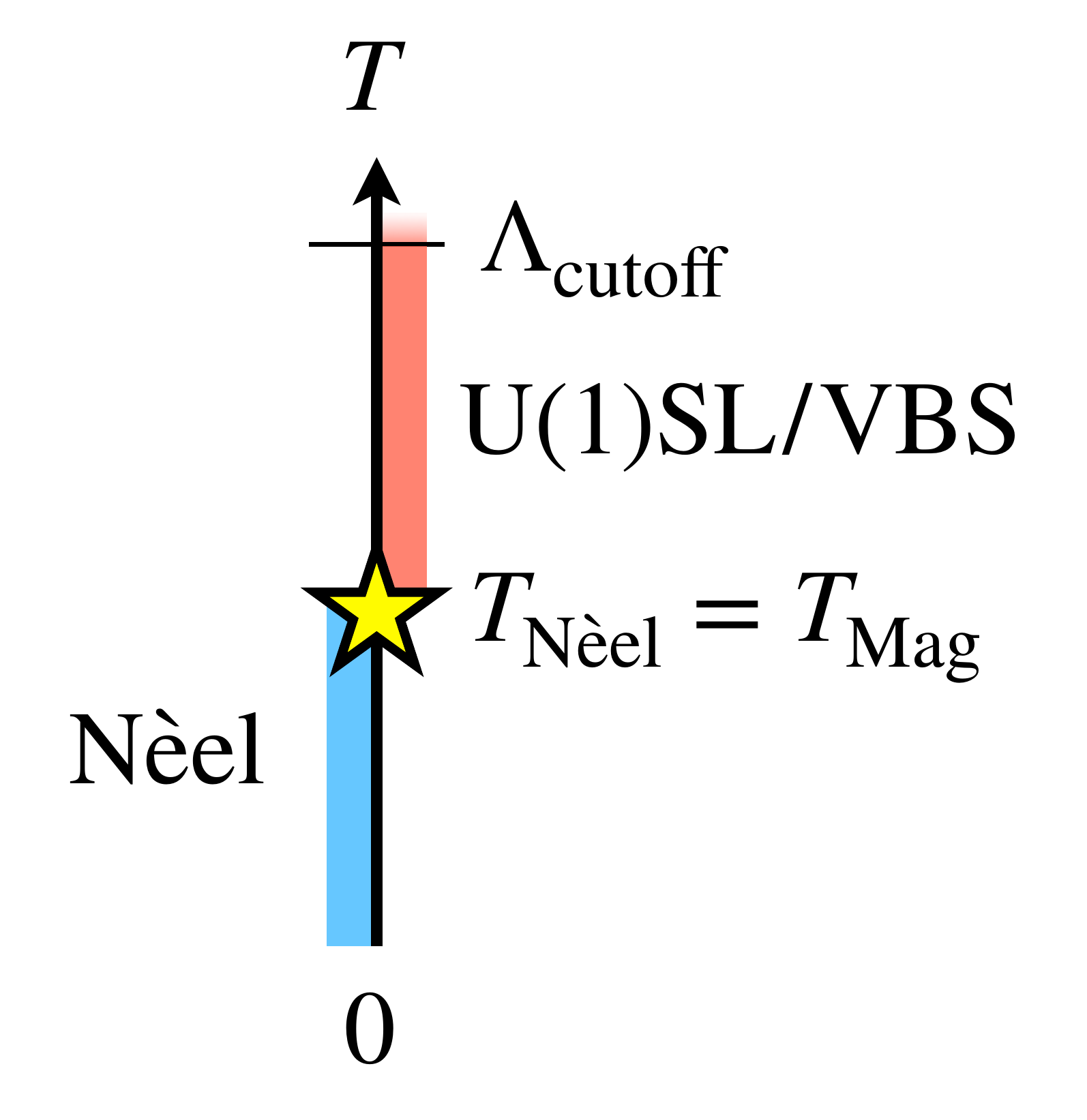}\label{fig: crit}}
 \subfloat[]{\includegraphics[scale=.27]{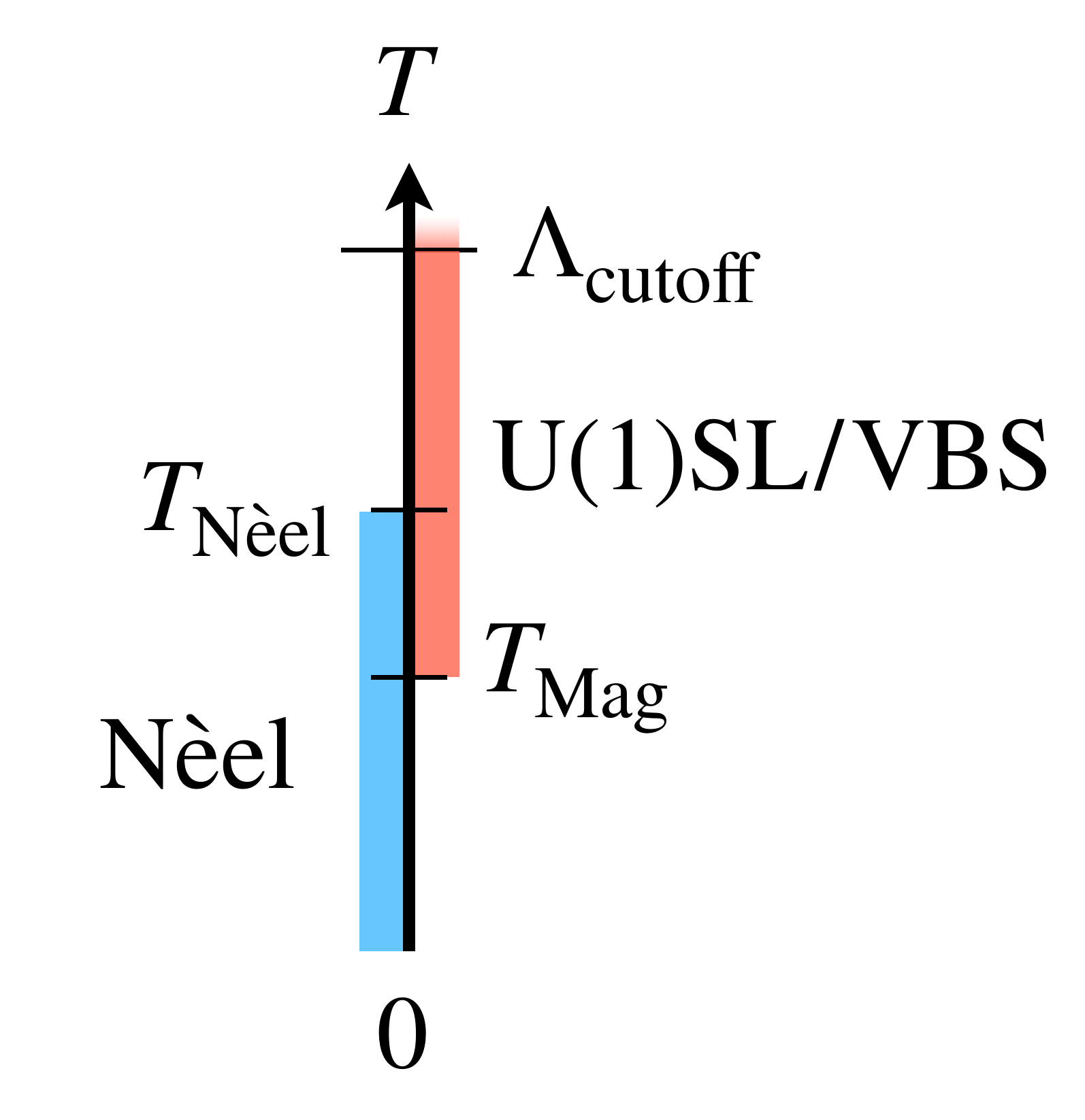}\label{fig: neel u1sl}}%
 \caption{
Possible scenarios for the finite temperature
phase diagram consistent with the anomalies
 In the first two examples, the N\'eel and $\U(1)$ spin liquid (SL)/VBS phases persist up to $T \sim \Lambda_\mathrm{cutoff}$,
while the last two show the cases with 
$ T_\mathrm{N\acute{e}el} = T_\mathrm{Mag}$
and $T_\mathrm{N\acute{e}el} > T_\mathrm{Mag}$, respectively.}
\end{figure}

In the simplest scenarios, the N\'eel and $\U(1)$ spin liquid/VBS phases persist up to the temperature $T \sim \Lambda_\mathrm{cutoff}$ as shown in Figs.~\ref{fig: neel} and~\ref{fig: u1sl}.
When $T > \Lambda_\mathrm{cutoff}$, the system could undergo a transition to a trivial gapped phase because the description by the $\CP^{N-1}$ model is no longer valid above the cutoff scale.

We shall consider a more nontrivial scenario.
First, suppose the model be in the N\'eel phase at zero temperature 
and the N\'eel order be destroyed at some temperature below the cutoff.
We define that temperature as $T_\mathrm{N\acute{e}el}$.
After the transition, the model must be in the $\U(1)$ spin liquid/VBS phase because the anomalies forbid a trivial gapped phase.
We introduce the temperature $T_\mathrm{Mag}$, at which the model becomes the $\U(1)$ spin liquid/VBS phase%
\footnote{Note that
$T_\mathrm{N\acute{e}el}$ and $T_\mathrm{Mag}$ are analogous to the temperatures for $\mC\mP$ symmetry breaking and deconfinement in the pure Yang-Mills theory at $\theta = \pi$, respectively~\cite{Gaiotto:2017yup}.
See also similar discussions for other gauge theories~\cite{Shimizu:2017asf, Komargodski:2017dmc, Tanizaki:2017mtm, Yonekura:2019vyz}.
}.
In general, $T_\mathrm{Mag}$ can be different from $T_\mathrm{N\acute{e}el}$,
but the anomalies yield the important constraint for the transition temperatures:
\begin{align}~\label{eq: inequality}
 T_\mathrm{N\acute{e}el} \ge T_\mathrm{Mag}.
\end{align}
This is because 
our simplified assumption on the possible phases does not allow a temperature window $T_\mathrm{N\acute{e}el} < T < T_\mathrm{Mag}$, between which a trivial phase could appear%
\footnote{
Nevertheless, it should be emphasized that the anomaly matching itself
\textit{does} allow $T_{\mathrm{N\acute{e}el}} < T_\mathrm{\U(1)SL}$ 
if the system shows an exotic phase matching all the 't Hooft anomalies 
in the temperature window $T_{\mathrm{N\acute{e}el}} < T < T_\mathrm{\U(1)SL}$.
It is interesting to investigate such an exotic scenario but beyond the scope of this paper.
}.
This scenario is schematically shown in Fig.~\ref{fig: neel u1sl}.
In a special case where the equality in Eq.~\eqref{eq: inequality} holds, 
the system may develop a critical behavior analogous to the deconfined quantum criticality~\cite{Senthil:2004-1, Senthil:2004-2, Senthil:2005}, 
or show a first-order phase transition (see Fig.~\ref{fig: crit}). 

\subsubsection{Large-$N$ phase diagram of $\CP^{N-1}$ nonlinear sigma model}

In order to demonstrate the actual realization of the 't Hooft anomaly, 
we here consider the finite-temperature $\CP^{N-1}$ nonlinear sigma model.
The genuine $\CP^{N-1}$ nonlinear sigma model is defined by the following Lagrangian
\begin{equation}
 \mL 
  = |D_a z|^2 + \lambda ( |z|^2 - N/g^2),
\end{equation}
where $z \equiv (z_1, \cdots, z_N)^t$ denotes a normalized $N$-component complex scalar field, whose normalization constraint $|z|^2 = N/g^2$ is imposed by the auxiliary field (Lagrange multiplier) $\lambda$.
Note that the dimensionful coupling $g^2$ is fixed in the large-$N$ limit.
Though we do not put the Maxwell term, the model enjoys the same symmetries as our $\CP^{N-1}$ linear model and also suffers from the same 't Hooft anomalies.
Thus, the anomaly matching argument presented above can also be applied to this model.

Let us then show that the large-$N$ phase diagram indeed realizes the plausible scenarios given in the previous subsection. 
For that purpose, we use the effective action in the leading large-$N$ expansion given by
\begin{equation}
 \Gamma [z,\lambda,a] = \int_0^\beta \diff^4 x 
 \left[ |D_a z|^2 + \lambda (|z|^2 - N/g^2)   \right]
 + N \mathrm{Tr} \log (-D_a^2 - \lambda).
\end{equation}
Then, 
assuming the homogeneous values for $\lambda(x) = \lambda_0$, $z(x) = z_0 = (\sqrt{N} v_0 ,0,\cdots, 0)$ and $a = 0$, we obtain the simplified expression for the effective potential $V (v_0,\lambda_0) \equiv \Gamma [z_0,\lambda_0,a=0]/(N \beta V)$ as
\begin{equation}
 V (v_0,\lambda_0) 
  = \lambda_0 ( v^2_0 - 1/g^2)  
  + \int^{\Lambda} \frac{\diff^3 k}{(2\pi)^3} 
  \left[   
   \frac{\omega (\bm{k})}{2} + T \log (1 - \rme^{-\omega(\bm{k})/T})
  \right],
  \label{eq:Effective-Pot}
\end{equation}
with $ \omega(\bm{k}) \equiv \sqrt{\bm{k}^2 + \lambda_0 }$.
Note that we employed a $3$-momentum cutoff regularization so that the momentum integral is performed within $|\bm{k}|< \Lambda \equiv \Lambda_\mathrm{cutoff}$.
In the following, we rescale all the dimensionful quantities by the cutoff scale $\Lambda$ such that 
$\bar{g} \equiv g \Lambda,~\bar{v}_0 \equiv v_0/\Lambda^2,~
\bar{\lambda}_0 = \lambda_0/\Lambda^2,~\bar{T} \equiv T/\Lambda$.

\begin{figure}[htb]
 \centering
 \includegraphics[width=0.45\linewidth]{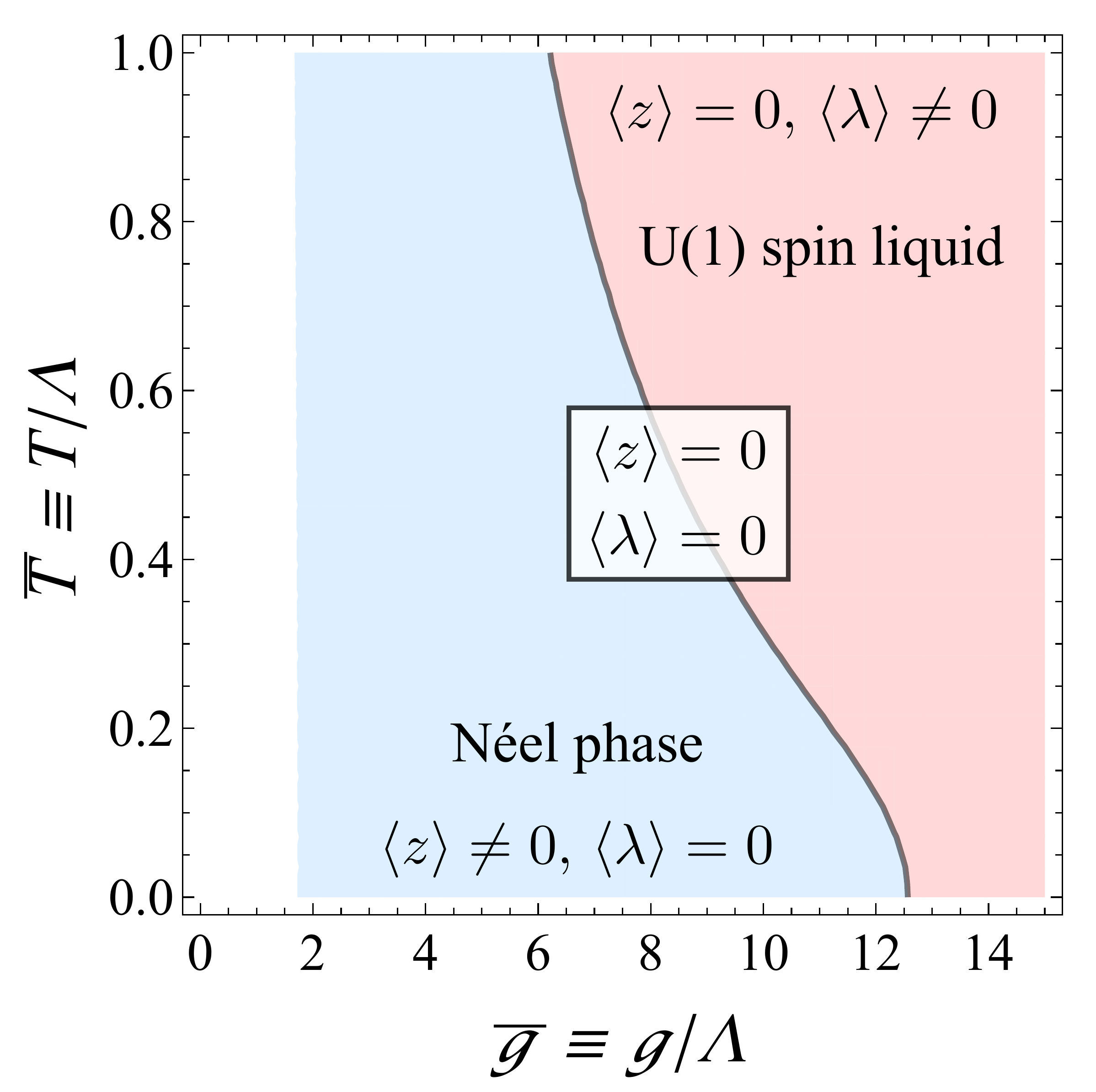}
 \caption{
 A large-$N$ phase diagram of the $\CP^{N-1}$ nonlinear sigma model.
 The N\'eel and $\U(1)$ spin liquid phases are separated by the critical line.
 }
\label{fig:4DCPN-phase}
\end{figure}

The resulting large-$N$ phase diagram from the effective potential \eqref{eq:Effective-Pot} is shown in Fig.~\ref{fig:4DCPN-phase}.
When we consider the zero-temperature limit, we find the critical coupling 
$\bar{g}_{\mathrm{cr}} \equiv 4 \pi$ separating the N\'eel and $\U(1)$ spin liquid phases. 
When $g > g_\mathrm{cr}$, $\lambda$ acquires a nonvanishing expectation value showing the $\U(1)$ spin liquid phase, and nonvanishing condensate $\braket{z}$ for $g < g_{\mathrm{cr}}$ indicates the N\'eel order. 
When one increases temperature, the value of the critical coupling decreases.
We emphasize that its phase structure fits in the scenarios discussed previously at any coupling $g$,
which shows the consistency with the anomaly matching at finite temperature%
\footnote{
In the $\U(1)$ spin liquid phase, we have an emergent gapless photon even though the original action does not contain the Maxwell term (see e.g., Ref.~\cite{Bando:1987br} for an interpretation of this emergent photon as a hidden local gauge boson). 
}.

\section{Summary and discussion}~\label{sec: summary}
In this paper, we have specified the 't Hooft anomalies in the $\CP^{N-1}$ model (or $N$-flavor abelian Higgs model) in $D$ spacetime dimensions with $D\ge3$ and applied them to constrain its low-energy dynamics.
Our investigation has revealed that the $\CP^{N-1}$ model possesses the two mixed 't Hooft anomalies: one between the $\PSU(N)_F$ flavor and $\U(1)_M^{[D-3]}$ magnetic symmetries, and the other between the $\U(1)_M^{[D-3]}$ magnetic and $\mR$ reflection symmetries.
We have also clarified the condition for these anomalies to survive in the presence of the charge-$n$ monopole object, which breaks the $\U(1)_M^{[D-3]}$ symmetry down to its discrete subgroup $(\Z_n^{[D-3]})_M$ (see Tables.~\ref{tbl: anom wo monopole} and \ref{tbl: anom w monopole} for a summary).
Thanks to the 't Hooft anomaly matching, the latter indicates that the $\CP^{N-1}$ model cannot have a unique gapped ground state even if one considers any $\PSU(N)_F$-breaking perturbations preserving the $\mR$ and $\U(1)_M^{[D-3]}$ symmetries.
For the case of $N=2$, 
typical perturbations are 
the staggered Zeeman magnetic field,
the easy-axis (easy-plane) potential,
and the Dzyaloshinskii-Moriya interaction:
\begin{equation}
V(n^\alpha) = 
 \sum_\alpha H^\alpha n^\alpha 
 + \sum_{\alpha} \mu^\alpha n^\alpha n^\alpha 
 % +\sum_{a \neq b} J^{ab} n^a n^b
 +\sum_{\alpha,\beta,\gamma} \kappa^\alpha_i \epsilon^{\alpha\beta\gamma} n^\beta
%D_\mu n^c.
 (\partial_i n^\gamma + (\kappa_i \times n)^\gamma ) .
\end{equation}
We list up the perturbations in Table~\ref{tbl: symmetry breaking} according to symmetry breaking patterns.
All the terms except the staggered Zeeman field preserve the magnetic symmetry and two of the reflection symmetries, and thus cannot result in a unique gapped ground state.
%%%%%%%%%%%%%%%%%%%%%
%%%		table
%%%%%%%%%%%%%%%%%%%%%
\begin{table}[t]
\centering
\begin{tabular}{c | c  c  c c c c c}
\hline \hline 
 \ & $\PSU(2)_F $  
     & $\mR_1 (\sim \mT)$ & $\mR_2 $  & $\mR_3$ & $\cdots$ & $\mR_D$ & $\U(1)^{[D-3]}_M$
\\ \hline
 $ H^{\alpha = x}$
 & $\ra \SO(2)_x$ & $\times$ & $\times$ & $\times$ & $\cdots$ & $\times$ & $\bigcirc$
\\ \hline
 $\mu^{\alpha = x}$
 & $\ra \mathrm{O}(2)_x$ & $\bigcirc$ & $\bigcirc$ & $\bigcirc$ & $\cdots $ & $\bigcirc$ & $\bigcirc$
\\ \hline
 $\kappa^{\alpha=x}_{i = 2}$
 & $\ra \mathrm{O}(2)_x$ & $\bigcirc$ & $\times$ & $\bigcirc$ & $\cdots$ & $\bigcirc$ & $\bigcirc$
\\
\hline \hline
\end{tabular}
 \caption{
 Symmetry properties of $\PSU(2)$-breaking perturbations 
for the $\CP^1$ model in the $D$ spacetime dimension.
Here
 $\bigcirc$ ($\times$) represents the perturbation preserves (completely breaks) the symmetry,
 and ``$\ra \SO(2)_x$'' (``$\ra \mathrm{O}(2)_x$'') means the flavor symmetry is broken to 
 its $\SO(2)_x$ ($ \mathrm{O}(2)_x$) subgroup generated by $\sigma^x$.
 } 
\label{tbl: symmetry breaking}
\end{table}
Furthermore, considering possible symmetry breaking scenarios, we have explicitly shown how our 't Hooft anomalies are saturated in the low-energy effective theories in the N\'eel, $\U(1)$ spin liquid, and valence bond solid phases. 
As an application of the anomaly matching, we have discussed the finite-temperature phase diagram of the $(3+1)$d $\CP^{N-1}$  model, where the anomalies involving $1$-form magnetic symmetry survives under the circle (thermal) compactification.

While we have specified the 't Hooft anomaly in the $\CP^{N-1}$ model---an effective field theoretical model of the $\SU(N)$ spin systems---it is an interesting issue to investigate its possible realization in the lattice model.
One may na\"ively expect that it could still survive in the underlying lattice model as an intrinsic (or a Lieb-Shultz-Mattis type) anomaly showing a projective representation of the corresponding symmetries.
If this is the case, the specified 't Hooft anomaly could restrict possible low-energy dynamics of the underlying lattice model beyond the regime where the field-theoretical description breaks down.
Nevertheless, this is not always the case, as shown in Ref.~\cite{Metlitski:2017fmd}; a so-called \textit{emergent anomaly} is present only in the field-theoretical model in sharp contrast to an intrinsic (Lieb-Shultz-Mattis type) one present even at the lattice scale.
This discrepancy is possible because we cannot trace the field-theoretical RG flow back to the UV lattice scale.
From the practical viewpoint to study the lattice model, the intrinsic anomaly is more robust and gives stronger constraints than the emergent anomaly.
Therefore, it is quite important to figure out a field theoretical 't Hooft anomaly belongs to which class of them, intrinsic or emergent.

We here discuss whether the $\mR \times (\Z_n)_M$ anomaly in the $(2+1)$d model is still present or not in an underlying spin model.
Since our $\mR \times (\Z_n)_M$ anomaly is its higher-dimensional generalization of the intrinsic $\mR \times \mC$ anomaly in the $2$d $\CP^{N-1}$ model studied in Ref.~\cite{Sulejmanpasic:2018upi}, it is natural to expect ours also belongs to the intrinsic one.
To clarify this, considering a lattice model with half-integer spins, we shall ask whether it is possible to place the spin on each site without spoiling the symmetries.
Note that a half-integer spin has the doubly-degenerated ground state when the time-reversal symmetry is respected (i.e., the Kramers degeneracy) due to the time-reversal anomaly.
To respect the time-reversal symmetry---corresponding to the reflection symmetry in our setup---we employ a construction using the Haldane spin chain: a $(1+1)$d SPT phase composed of integer spins and possessing the spin-$1/2$ at its edge.
Thanks to the cancellation between the bulk and edge contributions, the Haldane spin chain is free from the time-reversal anomaly.
The edge spin-$1/2$ degrees of freedom enable us to formulate a problem whether we can put the Haldane spin chains on $(2+1)$d lattice links appropriately without spoiling the $(\Z_n)_M$ lattice rotation symmetry. 
In this construction, the system has no time-reversal anomaly because the contributions from the links saturate the time-reversal anomalies on the sites.
However, corresponding to the $\mR \times (\Z_n)_M$ anomaly, the above time-reversal-respecting construction may not work if we impose the site-centered rotational symmetry%
\footnote{
This type of argument appears in Ref.~\cite{Po:2017, Shiozaki:2018yyj} as a generalization of the LSM theorem. See also Ref.~\cite{Song:2017} for related work.}.

\begin{figure}[ht]%
 \centering
 \subfloat[]{\includegraphics[scale=.3]{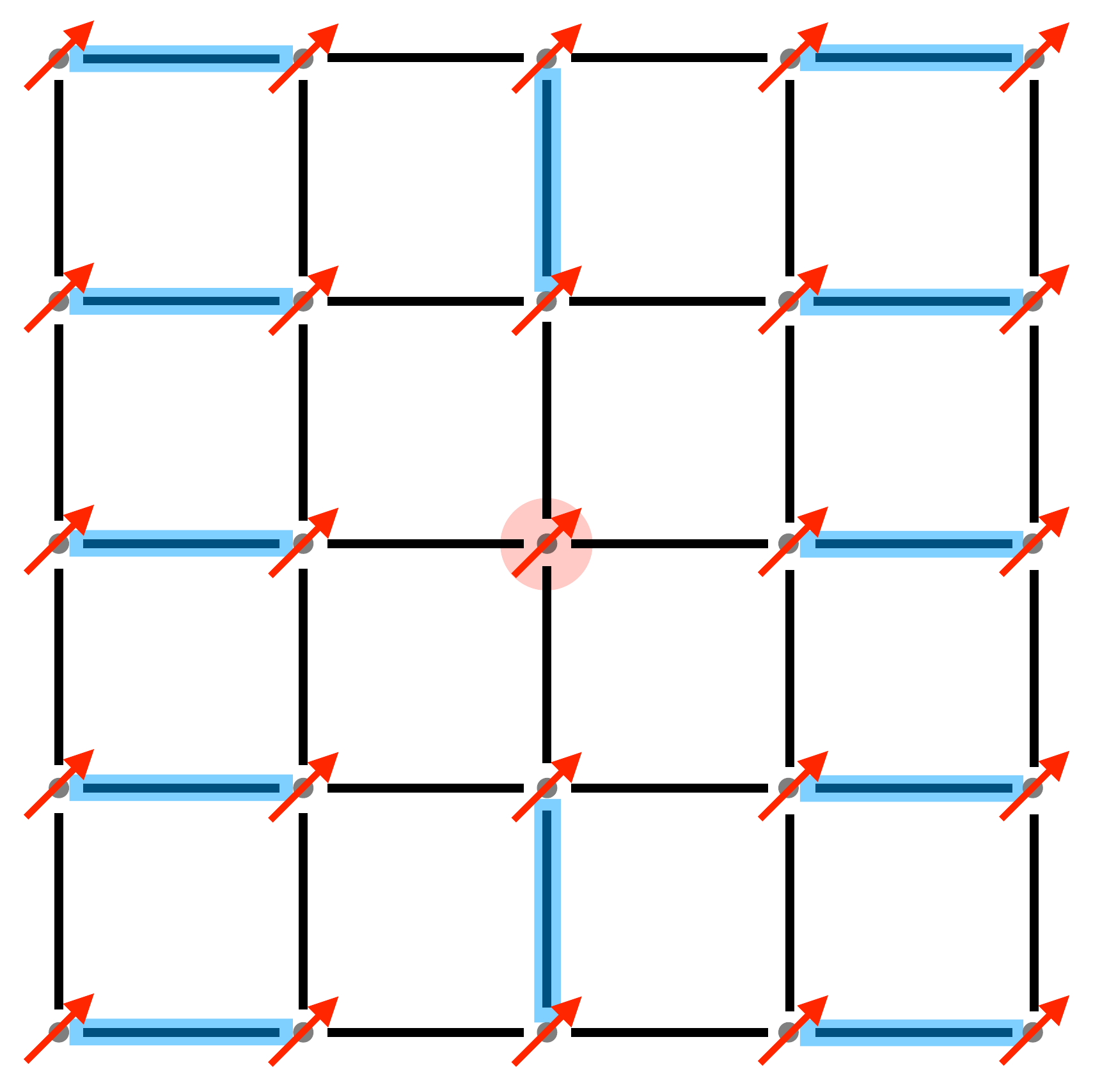}\label{fig: rectangular}}
 \hspace{50pt}
 \subfloat[]{\includegraphics[scale=.3]{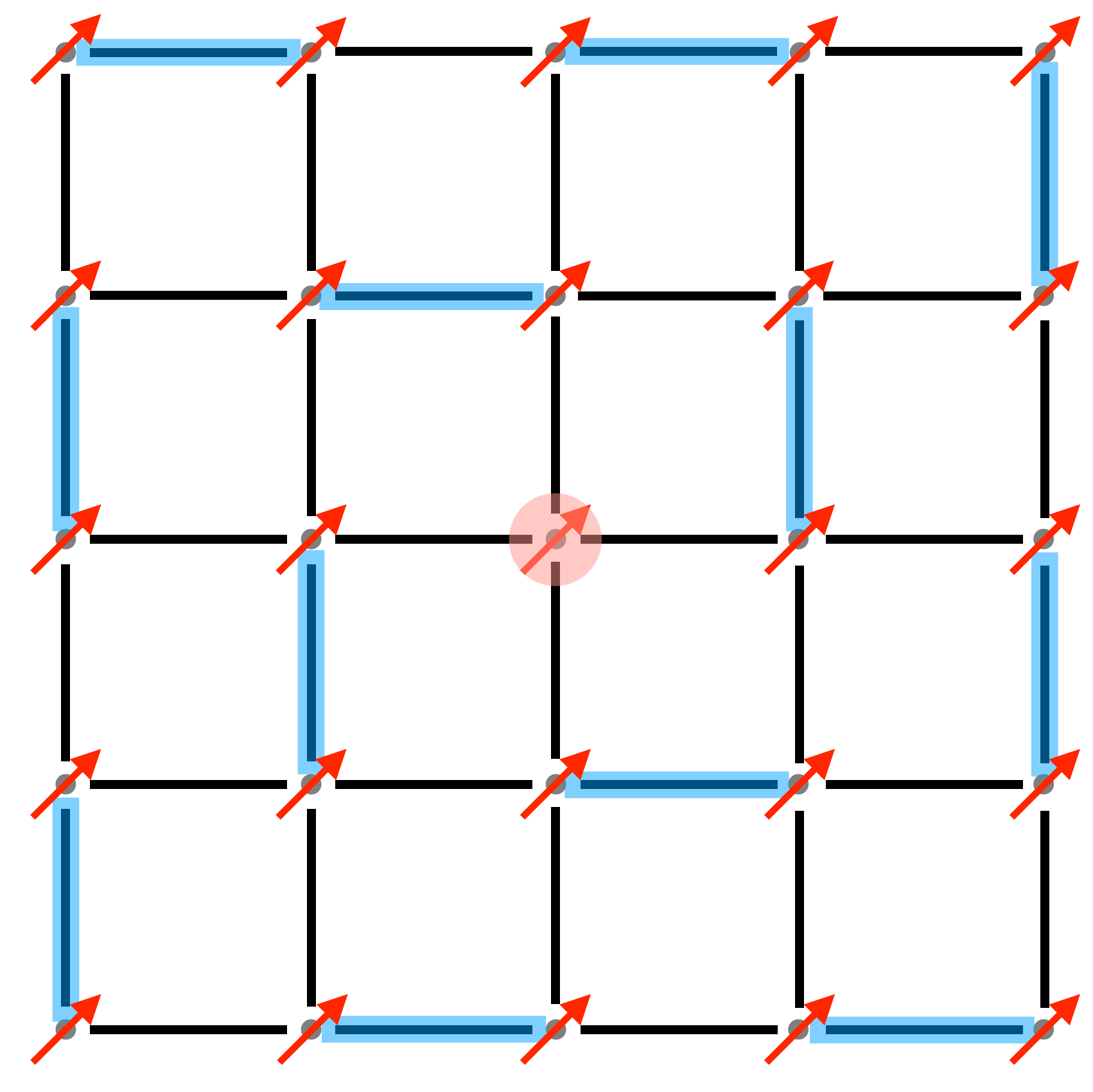}\label{fig: square}}\\
 \subfloat[]{\includegraphics[scale=.3]{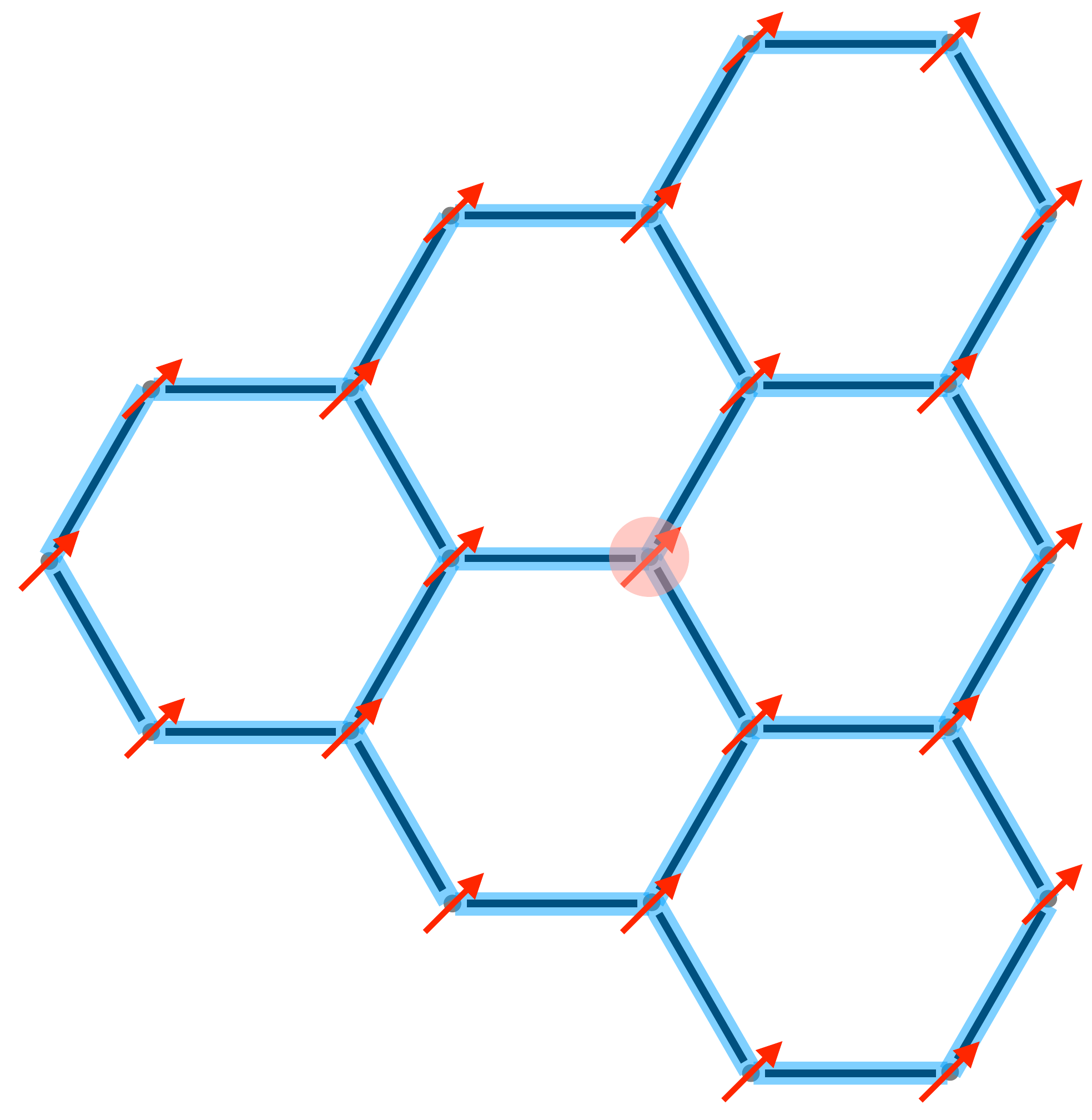}\label{fig: honeycomb}}%
 \caption{Construction of spin models on the (a) rectangular, (b) square, and (c) honeycomb lattices. A blue thick line represents the Haldane chain hosting a spin $1/2$ (a red arrow) at each edge.
 We never realize the spin at the rotation center (a red disk) as an edge of the Haldane chain
 on the first two lattices, while we can on the last lattice.}%
 \label{fig: lattices}%
\end{figure}

Let us consider a system on the rectangular lattice as a simple example.
In this case, we have to place the Haldane chains so that the system is invariant under $\pi$ rotation.
As shown in Fig.~\ref{fig: rectangular},
an isolated spin $1/2$ necessarily appears at the rotation center.
Thus, we never construct the spin $1/2$ system on the rectangular lattice by placing the Haldane chains respecting the rotational symmetry.
Likewise, such an isolated spin $1/2$ must exist, and the system is anomalous on the square lattice (see Fig.~\ref{fig: square}).
On the other hand, the Haldane chains completely saturate it, and no anomaly exists on the honeycomb lattice, as shown in Fig.~\ref{fig: honeycomb}.
This is because the time-reversal anomaly is a $\Z_2$ anomaly, and an odd number of one-half spins are anomalous.
This observation is also consistent with the construction of a trivial gapped ground state in a spin-$1/2$ system on the honeycomb lattice~\cite{Jian:2015,Kim:2016}.
All these results are consistent with Table.~\ref{tbl: anom w monopole},
implying the presence of the $\mR \times ( \Z_n )_M$ anomaly at the lattice scale.
We thus conclude that our $\mR \times ( \Z_n )_M$ anomaly is not an emergent anomaly but an  intrinsic one.

Although we restrict ourselves to the $\mR \times (\Z_n)_M$ anomaly in the $(2+1)$d model above, it is also interesting to ask whether and how 't Hooft anomalies in the $(3+1)$d model are present in the corresponding lattice model.
Since the magnetic symmetry in the $(3+1)$d $\CP^{N-1}$ model is a $1$-form symmetry, there could be such a discrete $1$-form symmetry even in the lattice model, and a related Lieb-Shultz-Mattis type anomaly.
Its identification can yield a further constraint on the finite-temperature phase diagram because the intrinsic anomaly prohibits the lattice model from being trivial beyond the cutoff $\Lambda_\mathrm{cutoff}$.
% \bsout{The identification of this may give us a lattice model showing higher-dimensional generalization of the competing order and possible exotic critical point between them, whose finite-temperature counterpart is most likely regarded as the $(2+1)$d deconfined quantum critical point. }
Furthermore, taking account of the persistence of $(3+1)$d 't Hooft anomalies at finite temperature, one can also ask whether there is an associated transport phenomenon or not.
Recent developments of quantum field theory in local thermal equilibrium enable us to investigate this~\cite{Banerjee:2012iz,Jensen:2012jh,Golkar:2015oxw,Chowdhury:2016cmh,Landsteiner:2016led,Hongo:2016mqm,Glorioso:2017lcn,Hongo:2019rbd}, and it is interesting to pursue possible manifestation of the 't Hooft anomaly in transport phenomena.
These are left for future work.

\acknowledgments

The authors greatly thank Y. Tanizaki for insightful comments.
One of the authors (T.F.) also thanks
J.-Y. Chen, Y. Fuji, A. Furusaki, S. Furuya, Y. Horinouchi, 
Z. Komargodski, M. Metlitski, Y. Nishida, and S. Ryu for stimulating discussion.
T.F. was supported by International Research Center for Nanoscience and Quantum Physics, Tokyo Institute of Technology, and by RIKEN Junior Research Associate Program.
M. H. was supported by the U.S. Department of Energy, Office of Science, Office of Nuclear Physics under Award Number DE-FG0201ER41195.
This work was partially supported by the “Topological Materials Science” (No. JP15K21717) KAKENHI on Innovative Areas from JSPS of Japan, the Ministry of Education, Culture, Sports, Science, and Technology(MEXT)-Supported Program for the Strategic Research Foundation at Private Universities ``Topological Science'' (Grant No. S1511006), and the RIKEN iTHEMS Program, in particular, iTHEMS STAMP working group.

%%%%%%%%%%%%%%%%%%%%%%%%%
%%%%		Appendix
%%%%%%%%%%%%%%%%%%%%%%%%%
\appendix

\section{Derivation of Eq.~\eqref{eq:Omega}}
\label{sec:Omega}
We here provide a derivation of Eq.~\eqref{eq:Omega} as is discussed in Ref.~\cite{Sulejmanpasic:2018upi}.
For that purpose, let us look for a condition imposed on $\Omega$.
Applying the above transformation twice, we get
\begin{align}
z(x) &\ra \Omega \Omega^\ast z (x),
\\
z^\+ (x) &\ra z^\+ (x) (\Omega \Omega^\ast)^\+,
\\
a(x) &\ra a (x).
\end{align}
Since $\mR_{\mu}$ is a $\Z_2$ symmetry,
$\Omega \Omega^\ast$ must be an element of the $\SU(N)$ center and thus $\Omega \Omega^\ast = \rme^{2 \pi \rmi k/N} \I_N$ ($k = 0, \cdots, N-1$)%
\footnote{Note that we can choose $\Omega \in \U(N)$ more generally.
Then, $\Omega \Omega^\ast$ is allowed to be an element of the $\U(N)$ center,
$\Omega \Omega^\ast = e^{i \theta}$ with $\theta \in [0, 2 \pi)$.
However, taking its determinant yields
$e^{i N \theta} = 1$, coming back to the $\SU(N)$ case.}.
This condition leads to
$\Omega = \rme^{2 \pi \rmi k/N} \Omega^t$.
As a result of this equation and its transpose, we obtain $\Omega = (\rme^{2 \pi \rmi k/N})^2 \Omega$, which leads to $\rme^{2 \pi \rmi k/N} = \pm1$.
We thus have only $\Omega = +\Omega^t$ for odd $N$
but $\Omega = \pm \Omega^t$ for even $N$.

\bibliography{anomalymatching}

\end{document}